\documentstyle[11pt,amssymb,epsfig,psfig]{article}
\makeatletter
\@addtoreset{equation}{section}

\makeatother
\def\beq{\begin{equation}}
\def\eeq{\end{equation}}

\def\text#1{\mbox{\scriptsize #1}}

\begin{document}

\title{Scalar Casimir effect for $D$-dimensional
spherically symmetric Robin boundaries}
\author{ Aram A. Saharian\footnote{%
E-mail address: saharyan@server.physdep.r.am}} \maketitle

\begin{center}
{\em Department of Physics, Yerevan State University, \\[0pt]
1 Alex Manoogian Street, 375049 Yerevan, Armenia}\\[0pt]
\end{center}

\bigskip

The vacuum expectation values for the energy-momentum tensor of a
massive scalar field with general curvature coupling and obeying
the Robin boundary condition on spherically symmetric boundaries in
$D$-dimensional space are investigated. The expressions are
derived for the regularized vacuum energy density and radial and
azimuthal stress components (i) inside and outside a single
spherical surface and (ii) in the intermediate region between two
concentric spheres. A regularization procedure is carried out by
making use of the generalized Abel-Plana formula for the series over
zeros of cylinder functions. The asymptotic behavior of the vacuum
densities near the sphere and at large distances is investigated.
A decomposition of the Casimir energy into volumic and surface
parts is provided for both cases (i) and (ii). We show that the
mode sum energy, evaluated as a sum of the zero-point energy
for each normal mode of frequency, and the volume integral of the
energy density in general are different, and argue that this
difference is due to the existence of an additional surface
energy contribution.

\bigskip

PACS number(s): 11.10.Kk, 11.10.Jj, 12.39.Ba

\bigskip

\section{Introduction}

Historically, the investigation of the Casimir effect (for a general
introduction, see \cite{Mostepanenko,Plunien,Milrev}) for a
perfectly conducting spherical shell was motivated by the Casimir
semiclassical model of an electron. In this model Casimir
suggested that Poincare stress, to stabilize the charged
particle, could arise from vacuum quantum fluctuations and the
fine structure constant can be determined by a balance between
the Casimir force (assumed attractive) and the Coulomb repulsion.
However, as has been shown by Boyer \cite{Boyer}, the Casimir
energy for the sphere is positive, implying a repulsive force.
This result was later reconsidered by a number of authors
\cite{DaviesSph,Balian,MiltonSph}. More recently new methods have
been developed for this problem including direct mode summation
techniques based on the zeta function regularization scheme
\cite{Blau}-\cite{Elizod} (for similar considerations in the
case of a dielectric ball see, for instance, the references given
in \cite{Nestdil}).

Investigation of the dimensional dependence of physical
quantities in the Casimir effect is of considerable interest. In
particular, for periodic boundary conditions this is
motivated by the idea of using the Casimir effect as a source for
dynamical compactification of the extra dimensions in
Kaluza-Klein models. In \cite{Ambjorn} the Casimir energy is
derived in a general hypercuboidal region and for various types
of boundary condition and field. For a spherical shell the
Casimir effect in an arbitrary number of dimensions is analyzed
in \cite{MiltonSc,MiltonVec} for a massless scalar field satisfying
Dirichlet and a special type of Robin (corresponding to the
electromagnetic TM modes) boundary conditions using the Green's
function method (see also \cite{Mildim}) and in \cite{Cognola}
for the electromagnetic field and massless scalar and spinor fields
(for the 4D fermionic Casimir effect see
\cite{Milfer,Baacke,Elizmas}) with various boundary conditions on
the basis of the zeta regularization technique.

However, the most of the previous studies on spherical
geometry were focused on global quantities such as the total energy
and the force acting on a shell. Investigation of the energy
distribution inside a perfectly reflecting spherical shell was
made in \cite{Olaussen1} in the case of QED and in
\cite{Olaussen2} for QCD. The distribution of the other
components for the energy-momentum tensor of the electromagnetic
field inside as well as outside the shell can be obtained from
the results of \cite {Brevik1,Brevik2}. In these papers
the consideration was carried out in terms of the Green's function
(for this formalism see \cite{MiltonSph}). Another efficient
method for extracting finite parts from the vacuum expectation values
of the local physical observables in the case of plane boundaries
is based on the Abel-Plana summation formula (see, e.g.,
\cite{Mostepanenko}). We have generalized this formula to include
curved boundaries \cite{Sahmat,Sahrev}. In \cite{Grig1,Sah2shert}
(see also \cite{Sahrev}) the calculations of the regularized
vacuum expectation values for the electromagnetic energy-momentum
tensor inside and outside a perfectly conducting spherical shell
and in the region between two concentric spheres are based on the
generalized Abel-Plana formula.

In this paper the vacuum expectation values of the
energy-momentum tensor are investigated for a massive scalar
field with general curvature coupling parameter $\xi $,
satisfying the Robin boundary condition on spherically symmetric
boundaries in $D$-dimensional space. As special cases they
include the results for the Diriclet, Neumann, TM, and conformally
invariant Hawking boundary conditions. Robin type conditions also
appear in considerations of the vacuum effects for a confined
charged scalar field in external fields (see, for instance,
\cite{Ambjorn2}) and in quantum gravity \cite{EsposQG}. In
addition to describing the physical structure of the quantum
field at a given point, the energy-momentum tensor acts as the
source of gravity in the Einstein equations. It therefore plays an
important role in modeling a self-consistent dynamics involving
the gravitational field \cite{Birrel}. To the author's knowledge
all previous investigations on the spherical Casimir effect for
scalar fields were concerned mainly with global properties
(interior and exterior energies, vacuum stress on a sphere). The
massless scalar field with Dirichlet boundary condition is
considered in
\cite{BenderHays,RomPR,Leseduarte,MiltonSc,Nesterenko,Cognola}
and for Neumann and Robin boundary conditions in
\cite{Leseduarte,MiltonVec,Nesterenko,Cognola}. The case of the
massive scalar field is investigated in \cite{Bordmas} (for the
heat-kernel coefficients and determinants, see \cite{BorEK,DowkCQG}
and references therein). In \cite{Deutsch,Kennedy} asymptotic
expansions for the renormalized energy-momentum tensor are
developed near an arbitrary smooth boundary in the case of
conformally and minimally coupled 4D massless scalars with
Dirichlet, Neumann, and Robin boundary conditions. Here we
consider the scalar vacuum inside and outside a single spherical
shell, and in the intermediate region between two concentric
spherical shells. For the latter geometry the general case is
investigated when the constants in the Robin boundary condition
are different for inner and outer spheres. To evaluate the
corresponding field products we use the mode sum method in
combination with the summation formulas from Ref. \cite{Sahmat}
(see also \cite{Sahrev}). For scalars with general curvature
coupling the essential point is the relation between the mode sum
energy, evaluated as a renormalized sum of the zero-point
energies for each normal mode of frequency, and the volume
integral of the renormalised energy density. For flat spacetime
backgrounds the first quantity does not depend on $\xi$, since
the normal modes are the same for fields with different
values of this parameter (for instance, for minimal and conformal
couplings). Nevertheless, the corresponding energy-momentum tensor
depends on $\xi$ and, in general, this is the case for the vacuum
energy distribution and hence for the integrated vacuum energy.
As a result, as was mentioned in \cite{Deutsch}, the mode
sum energy and integrated energy in general are different. (Note
that in most of the papers referred to above the first
quantity is considered.) Below for the geometries under
consideration we calculate both these quantities and argue that
this difference is due to the existence of an additional surface
energy contribution to the total vacuum energy, and the Casimir
energy decomposition into volume and surface parts is provided
(for a similar consideration in the case of parallel plate
geometry see \cite{RomeoSah}).

We have organized the paper as follows. In the next section we
consider the vacuum inside a sphere and derive formulas for
expectation values of the energy density and stresses. Section
\ref{sec:inen} is devoted to the corresponding global quantities,
such as the total interior Casimir energy and vacuum force acting on a
sphere. We show that the interior vacuum energy contains two
parts: volume and surface ones. The vacuum expectation values of
the energy-momentum tensor (EMT) for the region outside a sphere are
considered in section \ref{sec:outdens}. The expressions for the
total energy and force acting on a sphere from outside are
derived. The total outside Casimir energy is decomposed into
surface and volume parts. Further in this section a spherical
shell with zero thickness is considered and the corresponding
global quantities including interior and exterior parts are
evaluated. In section \ref{sec:twosurfdens} we consider the
vacuum energy-momentum tensor between two concentric spheres and
section \ref{sec:twosphtotal} is devoted to the global quantities
in this region. Section \ref{sec:conclusion} concludes the main
results of the present paper. In the appendix  we derive 
the summation formula
over zeros of the Bessel functions combination on the basis of the
generalized Abel-Plana formula. The vacuum
expectation values in the region between two spheres contain this
type of series.

\section{Vacuum EMT inside a spherical shell}
\label{sec:indens}

Consider a real scalar field $\varphi $ with curvature coupling parameter $%
\xi $ in $D$-dimensional space satisfying the Robin boundary condition
\begin{equation}
\left( A_{1}+B_{1}n^{i}\nabla _{i}\right) \varphi (x)=0  \label{robcond}
\end{equation}
on the sphere, where $A_{1}$ and $B_{1}$ are constants, $n^{i}$ is the
unit inward normal to the sphere, and $\nabla _{i}$ is the covariant
derivative operator. Of course, all results in the following will
depend on the ratio $A_1/B_1$ only. However, to keep the
transition to the Dirichlet and Neumann cases transparent we will
use the form (\ref{robcond}). The corresponding field equation has
the form
\begin{equation}
\left( \nabla _{i}\nabla ^{i}+m^{2}+\xi R\right) \varphi =0,\quad \nabla
_{i}\nabla ^{i}=\frac{1}{\sqrt{-g}}\partial _{i}\left( \sqrt{-g}%
g^{ik}\partial _{k}\right) ,  \label{fieldeq}
\end{equation}
where $R$ is the scalar curvature for the background spacetime. The values $%
\xi =0$, and $\xi =\xi _{c}$ with $\xi _{c}\equiv (D-1)/4D$
correspond to the minimal and conformal couplings, respectively.
Here we will consider the case of flat spacetime. The
corresponding metric energy-momentum tensor is defined as
(see, e.g., \cite{Birrel})
\begin{equation}
T_{ik}=(1-2\xi )\partial _{i}\varphi \partial _{k}\varphi +(2\xi
-1/2)g_{ik}\partial ^{l}\varphi \partial _{l}\varphi -2\xi \varphi \nabla
_{i}\nabla _{k}\varphi +(1/2-2\xi )m^{2}g_{ik}\varphi ^{2}.  \label{EMT1}
\end{equation}
It can be seen that by using the field equation this expression
can also be presented in the form
\begin{equation}
T_{ik}=\partial _{i}\varphi \partial _{k}\varphi +\left[ \left( \xi -\frac{%
1}{4}\right) g_{ik}\nabla _{l}\nabla ^{l} -\xi \nabla _{i}\nabla
_{k}\right] \varphi ^{2}, \label{EMT2}
\end{equation}
and the corresponding trace is equal to
\begin{equation}
T_{i}^{i}=D(\xi -\xi _{c})\nabla _{i}\nabla ^{i} \varphi
^{2}+m^{2}\varphi ^{2}. \label{trace}
\end{equation}
By virtue of Eq.(\ref{EMT2}) for the vacuum expectation values
(VEV's) of the EMT we have
\begin{equation}
\langle 0|T_{ik}(x)|0\rangle =\lim_{x^{\prime }\rightarrow
x}\partial _{i}\partial _{k}^{\prime }\langle 0|\varphi
(x)\varphi (x^{\prime })|0\rangle +\left[ \left( \xi
-\frac{1}{4}\right) g_{ik}\nabla _{l}\nabla ^{l} -\xi \nabla
_{i}\nabla _{k}\right] \langle 0|\varphi ^{2}(x)|0\rangle ,
\label{vevEMT}
\end{equation}
where $|0\rangle $ is the amplitude for the vacuum state. Note
that the VEV $\langle 0|\varphi (x)\varphi (x^{\prime })|0
\rangle \equiv G^{+}(x,x^{\prime })$ is known as a positive
frequency Wightman function. In Eq.(\ref{vevEMT}) instead of this
function one can choose any other bilinear function of fields
such as the Hadamard function, Feynman's Green function, etc. The
regularized vacuum EMT does not depend on the specific choice.
The expectation values (\ref{vevEMT}) are divergent. They are
divergent in unbounded Minkowski spacetime as well. In a flat
spacetime the regularization is performed by subtracting from
Eq.(\ref{vevEMT}) the corresponding Minkowskian part:
\begin{equation}
\langle T_{ik}(x)\rangle _{{\rm reg}}=\langle 0|T_{ik}(x)|0\rangle -\langle
0_{M}|T_{ik}(x)|0_{M}\rangle =\hat{\theta}_{ik}\langle \varphi (x)\varphi
(x^{\prime })\rangle _{{\rm reg}},  \label{regEMT}
\end{equation}
where $|0_{M}\rangle $ denotes the amplitude for the Minkowski
vacuum and the form of the operator $\hat{\theta}_{ik}$ directly
follows from Eq.(\ref{vevEMT}). Therefore the finite difference
between two divergent terms in Eq.(\ref{regEMT}) can be obtained
from the corresponding difference between the Wightman functions,
by applying a certain second-order differential operator and
taking the coincidence limit. To derive the expression for the
regularized VEV of the field bilinear product we will use the
mode summation method. By expanding the field operator over
eigenfunctions and using the commutation rules one can see that
\begin{equation}
\langle 0|\varphi (x)\varphi (x^{\prime })|0\rangle =\sum_{\alpha }\varphi
_{\alpha }(x)\varphi _{\alpha }^{\ast }(x^{\prime }),  \label{fieldmodesum}
\end{equation}
where $\left\{ \varphi _{\alpha }(x),\varphi _{\alpha }^{\ast
}(x^{\prime })\right\} $ is a complete set of positive and
negative frequency solutions to the field equation
(\ref{fieldeq}), satisfying boundary condition (\ref{robcond}).

In accordance with the symmetry of the problem under
consideration we shall use hyperspherical polar coordinates
$(r,\vartheta ,\phi )\equiv (r,\theta _{1},\theta _{2},\ldots
\theta _{n},\phi )$, $n=D-2$, related to the rectangular ones
$(x_1,x_2,...,x_D)$ by (see, for instance, \cite{Erdelyi}, Section 11.1)
\begin{eqnarray}
\label{sphcoord}
x_1 & = & r\cos \theta _1,\quad , x_2=r\sin \theta _1 \cos \theta _2,
...,\quad x_n=r\sin \theta _1\sin \theta _2...\sin \theta _{n-1}\cos
\theta _n \\
x_{D-1} &=& r\sin \theta _1\sin \theta _2...\sin \theta _{n}\cos \phi ,
\quad x_D=r\sin \theta _1\sin \theta _2...\sin \theta _{n}\sin \phi ,
\nonumber
\end{eqnarray}
where $0\leq \theta _k \leq \pi$, $k=1,...,n$ and $0\leq \phi \leq 2\pi $.
In the hyperspherical coordinates for the region
inside the sphere the complete set of solutions to 
Eq.(\ref{fieldeq}), regular at the origin, has the form
\begin{equation}
\varphi _{\alpha }(x)=\beta _{\alpha }r^{-n/2}J_{\nu
}(r\sqrt{\omega ^{2}-m^{2}})Y(m_{k};\vartheta ,\phi )e^{-i\omega
t},\quad \nu =l+n/2,\,\,l=0,1,2,\ldots ,  \label{eigfunc}
\end{equation}
where $m_{k}=(m_{0}\equiv l,m_{1},\ldots ,m_{n})$, and $m_{1},m_{2},
\ldots ,m_{n}$ are integers such that
\begin{equation}
0\leq m_{n-1}\leq m_{n-2}\leq \cdots \leq m_{1}\leq l,\quad
-m_{n-1}\leq m_{n}\leq m_{n-1},  \label{numbvalues}
\end{equation}
$J_{\nu }(z)$ is the Bessel function, and $Y(m_{k};\vartheta ,\phi )$ is
the surface harmonic of degree $l$ (see \cite{Erdelyi}, Section 11.2).
This last can be
expressed through the Gegenbauer or ultraspherical polynomial $C_{p}^{q}(x)$
of degree $p$ and order $q$ as
\begin{equation}
Y(m_{k};\vartheta ,\phi )=e^{im_{n}\phi }\prod_{k=1}^{n}\left( \sin \theta
_{k}\right) ^{|m_{k}|}C_{m_{k-1}-|m_{k}|}^{|m_{k}|+n/2-k/2}\left( \sin
\theta _{k}\right) . \label{surfharm}
\end{equation}
The corresponding normalization integral is in the form
\begin{equation}
\int \left| Y(m_{k};\vartheta ,\phi )\right| ^{2}d\Omega =N(m_{k}).
\label{harmint}
\end{equation}
The explicit form for $N(m_{k})$ is given in \cite{Erdelyi}, Section 11.3,
and will not be necessary for the following considerations in this
paper. From the addition theorem \cite{Erdelyi}, Section 11.4, one has
\begin{equation}
\sum_{m_{k}}\frac{1}{N(m_{k})}Y(m_{k};\vartheta , \phi
)Y^{*}(m_{k};\vartheta ^{\prime },\phi ^{\prime
})=\frac{2l+n}{nS_{D}}C_{l}^{n/2}(\cos \theta ), \label{adtheorem}
\end{equation}
where $S_{D}=2\pi ^{D/2}/\Gamma (D/2)$ is the total area of the
surface of the unit sphere in $D$-dimensional space, $\theta $ is
the angle between directions $(\vartheta ,\phi )$ and $(\vartheta
^{\prime },\phi ^{\prime })$, and sum is taken over the integer
values $m_{k},\,k=1,2,\ldots ,n$ in accordance with
Eq.(\ref{numbvalues}).

The coefficients $\beta _{\alpha }$ in Eq.(\ref{eigfunc}) can be 
found from the
normalization condition
\begin{equation}
\int \left| \varphi _{\alpha }(x)\right| ^{2}dV=\frac{1}{2\omega },
\label{normcond}
\end{equation}
where the integration goes over the region inside the sphere.
Substituting eigenfunctions (\ref{eigfunc}), and using the relation
(\ref{harmint}) for the spherical harmonics and the value for the
standard integral involving the square of the Bessel function, one
finds
\begin{equation}
\beta _{\alpha }^{2}=\frac{\lambda }{N(m_{k})\omega a}T_{\nu
}(\lambda a), \label{normcoef}
\end{equation}
with the notations
\begin{equation}
\lambda =\sqrt{\omega ^{2}-m^{2}},\quad T_{\nu
}(z)=\frac{z}{(z^{2}-\nu ^{2})J_{\nu }^{2}(z)+z^{2}J_{\nu
}^{'2}(z)}. \label{teka}
\end{equation}

From boundary condition (\ref{robcond}) on the sphere surface for
eigenfunctions (\ref{eigfunc}) one sees that the possible values
for the frequency have to be solutions to the following equation
\begin{equation}
AJ_{\nu }(z)+BzJ_{\nu }^{\prime }(z)=0,\quad z=\lambda a, \quad
A=A_{1}+ B_{1}n/2a,\quad B=-B_{1}/a.  \label{eigenmodes}
\end{equation}
It is well known (see, e.g., \cite{Watson,abramowiz}) that for real $A$, $B$ and $%
\nu >-1$ all roots of this equation are simple and real, except the case $%
A/B<-\nu $ when there are two purely imaginary zeros. Let us denote by $%
\lambda _{\nu ,k},\,k=1,2,\ldots ,$ the zeros of the function
$AJ_{\nu }(z)+BzJ_{\nu }^{\prime }(z)$ in the right half plane,
arranged in ascending order of the real part. Note that for the
Neumann boundary condition $A/B=1-D/2$, for the TE and TM
electromagnetic modes $B=0$ (Dirichlet) and $A/B=D/2-1$,
respectively, and for the conformally invariant Hawking boundary
condition \cite{Kennedy} one has $A/B=1/D-D/2$.

Substituting Eq.(\ref{eigfunc}) into Eq.(\ref {fieldmodesum}) and using
addition formula (\ref{adtheorem}) for the spherical harmonics, one
obtains
\begin{eqnarray}
\langle 0|\varphi (x)\varphi (x^{\prime })|0\rangle
&=&\frac{(rr^{\prime })^{-n/2}}{naS_{D}}\sum_{l=0}^{\infty
}(2l+n)C_{l}^{n/2}(\cos \theta )
\label{fieldmodesum1} \\
&&\times \sum_{k=1}^{\infty }\frac{\lambda _{\nu ,k}T_{\nu
}(\lambda _{\nu ,k})}{\sqrt{\lambda _{\nu ,k}^{2}+m^{2}a^2}}J_{\nu
}(\lambda _{\nu ,k}r/a)J_{\nu }(\lambda _{\nu ,k}r^{\prime
}/a)e^{i\sqrt{\lambda _{\nu ,k}^{2}/a^{2}+m^{2}}(t^{\prime
}-t)}.  \nonumber
\end{eqnarray}
To sum over $k$ we will use the generalized Abel-Plana
summation formula \cite{Sahmat,Sahrev}
\begin{eqnarray}
2\sum_{k=1}^{\infty }T_{\nu }(\lambda _{\nu ,k})f(\lambda _{\nu ,k})
&=&\int_{0}^{\infty }f(x)dx+\frac{\pi }{2}{\rm Res}_{z=0}f(z)\frac{\bar{Y}%
_{\nu }(z)}{\bar{J}_{\nu }(z)} {}  \nonumber \\
&-&\frac{1}{\pi }\int_{0}^{\infty }\ dx\frac{\bar{K}_{\nu }(x)}{\bar{I}_{\nu
}(x)}\left[ e^{-\nu \pi i}f(xe^{\pi i/2})+e^{\nu \pi i}f(xe^{-\pi i/2})%
\right] ,  \label{sumJ1anal}
\end{eqnarray}
where, following \cite{Sahmat}, for a given function $F(z)$ we use the
notation
\begin{equation}
\bar{F}(z)\equiv AF(z)+BzF^{\prime }(z).  \label{barnot}
\end{equation}
Formula (\ref{sumJ1anal}) is valid for functions $f(z)$
analytic in the right half plane and satisfying the conditions
\begin{equation}
f(ze^{\pi i})=-e^{2\nu \pi i}f(z)+o(z^{2\nu -1}),\quad z\rightarrow 0,
\label{cond1}
\end{equation}
\begin{equation}
|f(z)|<\varepsilon (x)e^{c|y|},\,\,c<2,\,\,\varepsilon (x)\rightarrow 0,\quad
x\rightarrow \infty .  \label{cond2}
\end{equation}
This formula can be generalized in the case of the existence of
purely imaginary zeros for the function $\bar{J}_{\nu }(z)$ by adding
the corresponding residue term and taking the principal value of
the integral on the right (see \cite{Sahrev}). However, in the
following we will assume values of $A/B$ for which all these
zeros are real.

As the function $f(z)$ in Eq.(\ref{sumJ1anal}) let us choose
\begin{equation}
f(z)=\frac{z}{\sqrt{z^{2}+m^{2}a^{2}}}J_{\nu }(zr/a)J_{\nu }(zr^{\prime
}/a)e^{i\sqrt{z^{2}/a^{2}+m^{2}}(t^{\prime }-t)}.  \label{fAPF}
\end{equation}
This function has branch points on the imaginary axis. As has
been shown in \cite{Sahmat,Sahrev} the formula (\ref{sumJ1anal})
can be used for functions having this type of branch point
as well. As $f(z)\sim z^{2\nu +1}$, $z\rightarrow 0$ condition
(\ref{cond1}) is satisfied. Using the asymptotic formulae for the
Bessel function (see, e.g., \cite{abramowiz})
it is easy to see that condition (\ref{cond2}) is satisfied if $%
r+r^{\prime }+|t-t^{\prime }|<2a$. (Note that this condition is satisfied
in the coincidence limit $r=r'$, $t=t'$ for interior points, $r<a$.)
Assuming that this is the case and
applying to the sum over $k$ in Eq.(\ref{fieldmodesum}) formula (\ref
{sumJ1anal}) with $f(z)$ from Eq.(\ref{fAPF}) one obtains
\begin{eqnarray}
\langle 0|\varphi (x)\varphi (x^{\prime })|0\rangle &=&\frac{1}{2naS_{D}}%
\sum_{l=0}^{\infty }\frac{2l+n}{(rr^{\prime })^{n/2}}C_{l}^{n/2}(\cos \theta
)\left[ \int_{0}^{\infty }{f(z)dz}\right.   \label{unregWight} \\
&-&\left. \frac{2}{\pi }\int_{ma}^{\infty }dz\,z\frac{\bar{K}_{\nu }(z)}{%
\bar{I}_{\nu }(z)}\frac{I_{\nu }(zr/a)I_{\nu }(zr^{\prime }/a)}{\sqrt{%
z^{2}-m^{2}a^{2}}}\cosh \left[ \sqrt{z^{2}/a^{2}-m^{2}}(t^{\prime
}-t)\right] \right] ,  \nonumber
\end{eqnarray}
where we have used the result that the difference of the radicals
is nonzero above the branch point only and introduced the modified
Bessel functions. The contribution of the term in the first
integral to the VEV does not depend on
$a$, whereas the contribution of the second one tends to zero as $%
a\rightarrow \infty $. It follows from here that the first term is the
corresponding function for the unbounded Minkowski space:
\begin{equation}
\langle 0_{M}|\varphi (x)\varphi (x^{\prime })|0_{M}\rangle =\frac{1}{2nS_{D}%
}\sum_{l=0}^{\infty }\frac{2l+n}{(rr^{\prime })^{n/2}}C_{l}^{n/2}(\cos
\theta )\int_{0}^{\infty }dz\,\frac{ze^{i\sqrt{z^{2}+m^{2}}(t^{\prime }-t)}}{%
\sqrt{z^{2}+m^{2}}}J_{\nu }(zr)J_{\nu }(zr^{\prime }) . \label{Mink}
\end{equation}
This can be seen also by direct evaluation. Indeed, the sum over
$l$ can be summarized using the Gegenbauer addition theorem for
the Bessel function \cite{abramowiz}. The value of the remaining
integral involving the Bessel function can be found, e.g., in
\cite{Prudnikov}. As a result we obtain the standard expression
for the $D$-dimensional Minkowskian Wightman function:
\begin{equation}
\langle 0_{M}|\varphi (x)\varphi (x^{\prime })|0_{M}\rangle =\frac{m^{(D-1)/2}%
}{(2\pi )^{(D+1)/2}}\frac{K_{(D-1)/2}\left( m\sqrt{({\bf x}-{{\bf x^{\prime }}}%
)^{2}-(t-t^{\prime })^{2}+i\varepsilon }\right) }{\left[ ({\bf x}-{{\bf %
x^{\prime }}})^{2}-(t-t^{\prime })^{2}+i\varepsilon \right]
^{(D-1)/4}}, \label{MinkWight}
\end{equation}
where $x=(t,{\bf x})$, $\varepsilon >0$ for $t>t^{\prime }$, and $\varepsilon
<0$ for $t<t^{\prime }$. Using the Feynman Green function in 
Eq.(\ref{vevEMT}) we would obtain the same expression (\ref{MinkWight})
with $\varepsilon >0$.

Hence we see that the application of the generalized Abel-Plana
formula allows us to extract from the bilinear field product the
contribution due to the unbounded Minkowski spacetime. Combining
Eqs.(\ref{fieldmodesum1}) and (\ref{Mink}) for the regularized
Wightmann function one obtains
\begin{eqnarray}
\langle \varphi (x)\varphi (x^{\prime })\rangle _{{\rm reg}} &=&\langle
0|\varphi (x)\varphi (x^{\prime })|0\rangle -\langle 0_{M}|\varphi
(x)\varphi (x^{\prime })|0_{M}\rangle =\frac{-1}{n\pi S_{D}}%
\sum_{l=0}^{\infty }\frac{2l+n}{(rr^{\prime })^{n/2}}C_{l}^{n/2}(\cos \theta
)  \nonumber \\
&\times &\int_{m}^{\infty }dz\,z\frac{\bar{K}_{\nu }(za)}{\bar{I}_{\nu }(za)}%
\frac{I_{\nu }(zr)I_{\nu }(zr^{\prime
})}{\sqrt{z^{2}-m^{2}}}\cosh \!\left[
\sqrt{z^{2}-m^{2}}(t^{\prime }-t)\right] .  \label{regWight}
\end{eqnarray}
Let us recall that the response of the particle detector at a given
state of motion is determined by this function
\cite{Birrel,Davdet}. Performing the limit $x^{\prime
}\rightarrow x$ and using the value
\begin{equation}
C_{l}^{n/2}(1)=\frac{\Gamma (l+n)}{\Gamma (n)l!}  \label{Cl1}
\end{equation}
(see \cite{Erdelyi}, Section 11.1) for the regularized
field square we get
\begin{equation}
\langle \varphi ^{2}(r)\rangle _{{\rm reg}}=-\frac{1}{\pi ar^{n}S_{D}}%
\sum_{l=0}^{\infty }D_{l}\int_{ma}^{\infty }dz\ \frac{\bar{K}_{\nu }(z)}{%
\bar{I}_{\nu }(z)}\frac{zI_{\nu
}^{2}(zr/a)}{\sqrt{z^{2}-m^{2}a^2}}, \label{regsquare}
\end{equation}
where
\begin{equation}\label{Dlang}
  D_{l}=(2l+D-2)\frac{\Gamma (l+D-2)}{\Gamma (D-1)\,l!}
\end{equation}
is the degeneracy of each angular mode with given $l$.

The VEV for the EMT can be evaluated by substituting
Eqs.(\ref{regWight}) and (\ref {regsquare}) into Eq.(\ref{regEMT}),
where the operator $\hat{\theta}_{ik}$ is defined in accordance
with Eq.(\ref{vevEMT}). From the symmetry of the problem under
consideration it follows that $\langle T_{ik}\rangle _{{\rm
reg}}$ is a combination of the second rank tensors constructed
from the metric $g_{ik}$, the unit vector $\hat{t}^{i}$ in the time
direction, and the unit vector $n^{i}$ in the radial direction. We will
present this combination in the form
\begin{equation}
\langle T_{i}^{k}\rangle _{{\rm reg}}=\varepsilon \hat{t}_{i}\hat{t}%
^{k}+pn_{i}n^{k}+p_{\perp }\left( \hat{t}_{i}\hat{t}^{k}-n_{i}n^{k}-\delta
_{i}^{k}\right) ={\rm diag}\left( \varepsilon ,-p,-p_{\perp },\ldots
,-p_{\perp }\right) ,  \label{diagEMT}
\end{equation}
where the vacuum energy density $\varepsilon $ and the effective pressures in
radial, $p$, and azimuthal, $p_{\perp }$, directions are functions of the
radial coordinate only. As a consequence of the continuity equation $\nabla
_{k}\langle T_{i}^{k}\rangle _{{\rm reg}}=0$, these functions are related by
the equation
\begin{equation}
r\frac{dp}{dr}+(D-1)(p-p_{\perp })=0.  \label{conteq}
\end{equation}
From Eqs.(\ref{regEMT}), (\ref{regWight}), and (\ref{regsquare}) for the EMT
components one obtains
\begin{equation}
q(a,r)=-\frac{1}{2\pi a^{3}r^{n}S_{D}}\sum_{l=0}^{\infty
}D_{l}\int_{ma }^{\infty }dz\,z^{3}\frac{\bar{K}_{\nu
}(z)}{\bar{I}_{\nu }(z)}\frac{F_{\nu }^{(q)}\left[ I_{\nu
}(zr/a)\right] }{\sqrt{z^{2}-m^{2}a^2}},\quad q=\varepsilon
,\,p,\,p_{\perp },\quad r<a,  \label{q1in}
\end{equation}
where for a given function $f(y)$ we have introduced the notations
\begin{eqnarray}
F_{\nu }^{(\varepsilon )}\left[ f(y)\right] &=&(1-4\xi )\left[ f^{^{\prime
}2}(y)-\frac{n}{y}f(y)f^{\prime }(y)+\left( \frac{\nu ^{2}}{y^{2}}-\frac{%
1+4\xi -2(mr/y)^{2}}{1-4\xi }\right) f^{2}(y)\right]  \label{Fineps} \\
F_{\nu }^{(p)}\left[ f(y)\right] &=&f^{^{\prime }2}(y)+\frac{\xi _{1}}{y}%
f(y)f^{\prime }(y)-\left( 1+\frac{\nu ^{2}+\xi
_{1}\frac{n}{2}}{y^{2}}\right)
f^{2}(y),\;\xi _{1}=4(n+1)\xi -n  \label{Finperad} \\
F_{\nu }^{(p_{\perp })}\left[ f(y)\right] &=&(4\xi -1)f^{^{\prime }2}(y)-%
\frac{\xi _{1}}{y}f(y)f_{\nu }^{\prime }(y)+\left[ 4\xi
-1+\frac{\nu ^{2}(1+\xi _{1})+\xi
_{1}\frac{n}{2}}{(n+1)y^{2}}\right] f^{2}(y) . \label{Finpeaz}
\end{eqnarray}
It can easily be seen that components (\ref{q1in}) satisfy
Eq.(\ref {conteq}) and are finite for $r<a$. The formulas
(\ref{q1in}) may be derived in another, equivalent way,
introducing into the divergent mode sum
\begin{equation}
\langle 0|T_{ik}\{ \varphi (x), \varphi (x)\} |0\rangle = \sum
_{\alpha }T_{ik}\{ \varphi _{\alpha }(x), \varphi _{\alpha }^{*}(x)\}
\label{EMTmodesum}
\end{equation}
a cutoff function and applying the summation formula
(\ref{sumJ1anal}). The latter allows one to extract the corresponding
Minkowskian part by a manifestly cutoff independent method.

\begin{figure}[tbph]
\begin{center}
\begin{tabular}{ccc}
\epsfig{figure=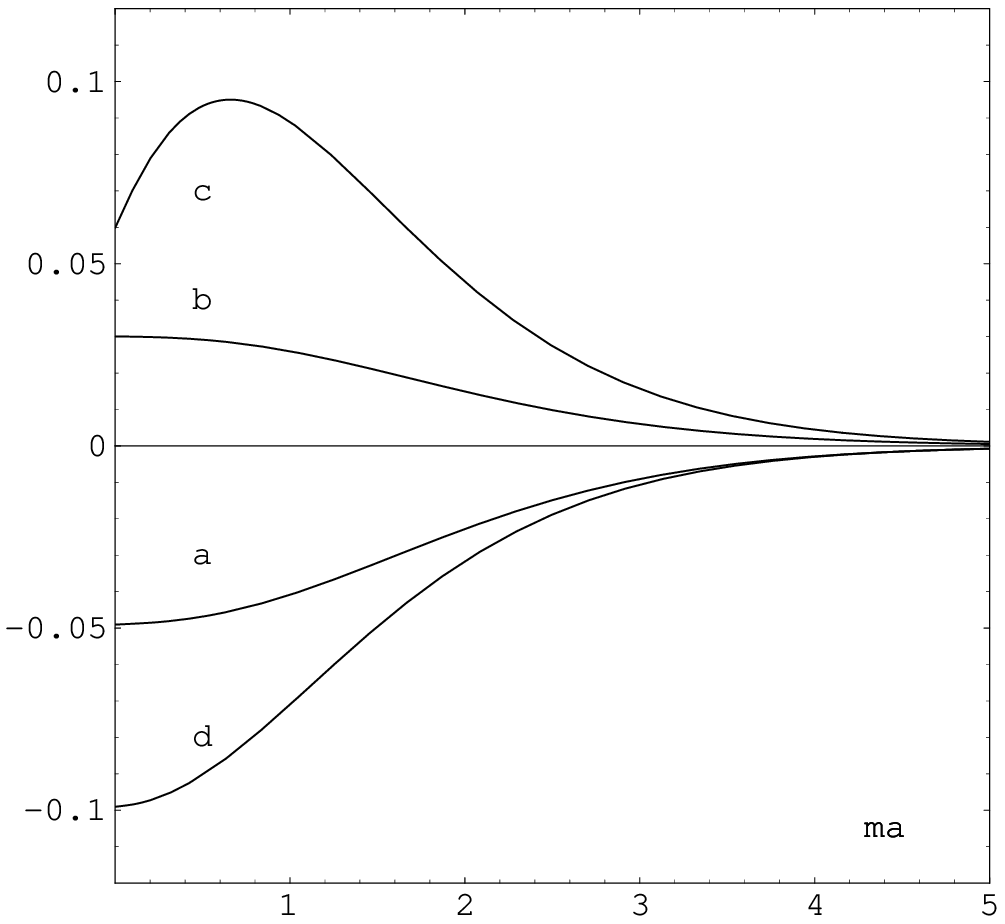,width=7cm,height=7cm} & \hspace*{0.5cm} & %
\epsfig{figure=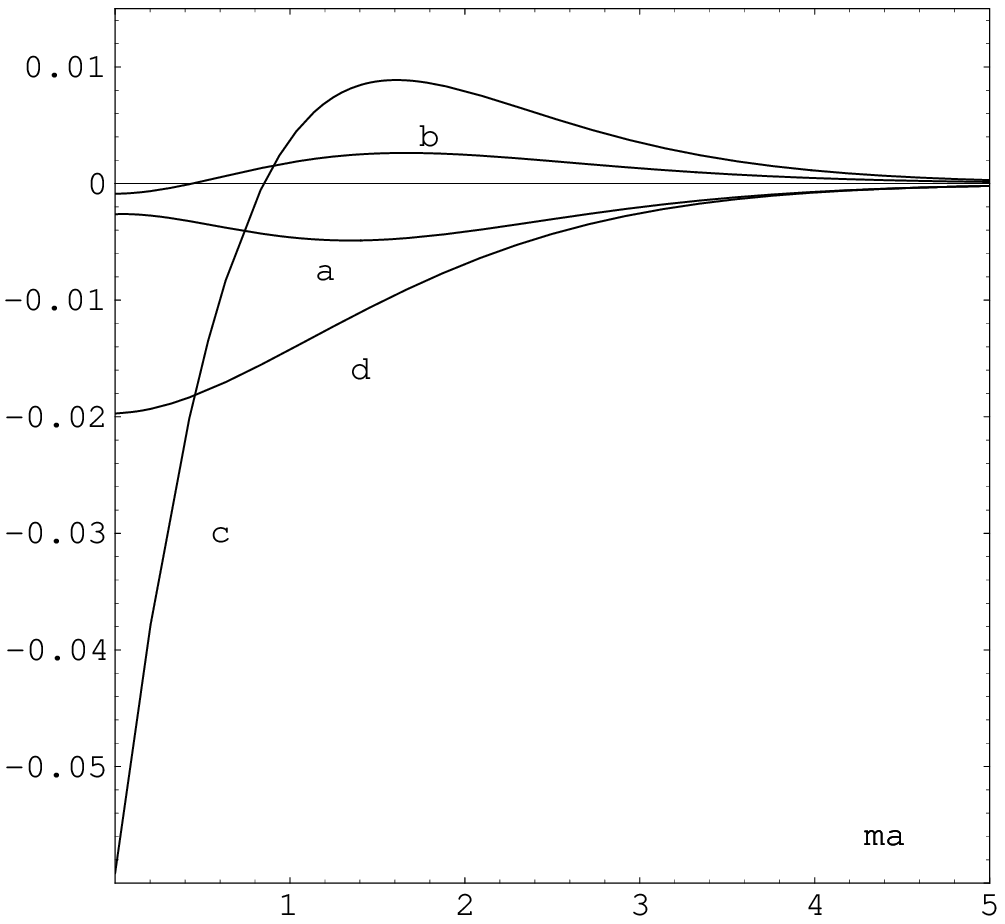,width=7cm,height=7cm}
\end{tabular}
\end{center}
\caption{ The Casimir energy density, $a^{D+1}\protect\varepsilon
$, (a, c) and vacuum pressure, $a^{D+1}p$, (b, d) at sphere center
for minimally (left) and conformally (right) coupled Dirichlet
(a, b) and Neumann (c, d) scalars in $D=3$ versus $ma$.}
\label{figspcent}
\end{figure}

At the sphere center the nonzero contribution to VEV (\ref
{q1in}) comes from the summands with $l=0$ and $l=1$ and one has
\begin{eqnarray}
\varepsilon (a,0) &=&\frac{1}{2^{D}\pi ^{D/2+1}\Gamma (D/2)
}\int_{m}^{\infty }\frac{z^{D+1}dz}{\sqrt{z^{2}-m^{2}}}
\label{encentre} \\
&&\times \left[ \left( 4\xi +1-2\frac{m^{2}}{z^{2}}\right) \frac{\bar{K}%
_{D/2-1}(az)}{\bar{I}_{D/2-1}(az)}+(4\xi -1)\frac{\bar{K}_{D/2}(az)%
}{\bar{I}_{D/2}(az)}\right] ,  \nonumber
\end{eqnarray}
\begin{eqnarray}
p(a,0) &=&p_{\perp }(a,0)=-\frac{1}{2^{D}\pi ^{D/2+1}D\Gamma (D/2)
}\int_{m}^{\infty }\frac{z^{D+1}dz}{\sqrt{z^{2}-m^{2}}}
\label{pcentre} \\
&&\times \left[ \left( \xi _{1}-2\right) \frac{\bar{K}_{D/2-1}(az)}{\bar{%
I}_{D/2-1}(az)}+\xi _{1}\frac{\bar{K}_{D/2}(az)}{\bar{I}%
_{D/2}(az)}\right] . \nonumber
\end{eqnarray}
Note that for the conformally coupled massless scalar
\begin{equation}
\varepsilon (a,0)=Dp(a,0).  \label{enpcentre}
\end{equation}
This can also be obtained directly from the zero trace condition.
The results of the corresponding numerical evaluation for the
Dirichlet and Neumann minimally and conformally coupled scalars
in $D=3$ are presented in Fig. \ref{figspcent}.
In the limit of large mass, $ma\gg 1$, using the asymptotic formulas for the
modified Bessel functions for large arguments and the value of the integral
\begin{equation}
\int_{\mu }^{\infty }\frac{z^{m+1}e^{-2z}}{\sqrt{z^{2}-\mu ^{2}}}%
dz=(-1)^{m}\mu ^{m+1}\frac{\partial ^{m}K_{1}(2\mu )}{\partial (2\mu )^{m}},
\label{integralforcent}
\end{equation}
from Eqs.(\ref{encentre}) and (\ref{pcentre}) one obtains
\begin{equation}
\varepsilon (a,0)\approx -\frac{D}{D-1}p(a,0)\approx \frac{(4\xi
-1)(am)^{D+1/2}e^{-2am}}{2^{D}\pi ^{(D-1)/2}\Gamma (D/2)
a^{D+1}}\left( 2\delta _{B0}-1\right) .  \label{centrelargemass}
\end{equation}

The expectation values (\ref{q1in}) diverge at the sphere surface,
$r\rightarrow a$ (note that for a given $l$ the integrals over
$z$ diverge as $(a-r)^{-2}$). The corresponding asymptotic
behavior can be found using the uniform asymptotic expansions for
the modified Bessel functions, and the leading terms have the form
\begin{eqnarray}
p &\sim &\frac{(D-1)\Gamma ((D+1)/2)(\xi -\xi _{c})}{2^{D}\pi
^{(D+1)/2}a(a-r)^{D}}\left( 1-2\delta _{B0}\right) , \label{peinasimp} \\
\varepsilon &\sim &-p_{\perp }\sim -\frac{D\Gamma ((D+1)/2)(\xi
-\xi _{c})}{2^{D}\pi ^{(D+1)/2}(a-r)^{D+1}}\left( 1-2\delta
_{B0}\right) . \label{epsinasimp}
\end{eqnarray}
These terms do not depend on mass or Robin coefficient $B_1$,
and have opposite signs for Dirichlet and Neumann boundary
conditions. Surface divergences in renormalized expectation
values for EMT are well known in quantum field theory with
boundaries \cite{Balian,Deutsch,Kennedy}. In the case of $D=3$
massless fields the corresponding asymptotic series near an
arbitrary smooth boundary are presented in
\cite{Deutsch,Kennedy}.  In particular, for $D=3$, $\xi =0$ from
Eqs.(\ref{peinasimp}) and (\ref{epsinasimp}) we obtain the leading
terms given in \cite {Deutsch} for the minimally coupled Dirichlet
and Neumann scalar fields. Taking the limit $a,r\rightarrow
\infty $, $a-r={\rm const}$, one obtains the leading terms for the
asymptotic behavior near the single plate (see, for instance,
\cite{RomeoSah}). For a conformally coupled scalar the
coefficients for the leading terms are zero and $\varepsilon
,p_{\perp }\sim (a-r)^{-D}$, $p\sim (a-r)^{1-D}$. In general
asymptotic series can be developed in powers of the distance from
the boundary. The corresponding subleading coefficients will
depend on the mass, Robin coefficient, and sphere radius.

In the case $D=1$ we obtain VEV's for the one - dimensional segment
$-a\leq x\leq a$. Due to the gamma function in the denominator of the
expression for $D_{l}$ in Eq.(\ref{Dlang}), now the only nonzero
coefficients are $D_{0}=D_{1}=1$. Using the standard expressions
for the functions $I_{\pm 1/2}(z)$ and $K_{\pm 1/2}(z)$ we can
easily see that $F_{\pm 1/2}^{(p)}\left[ I_{\pm 1/2}(z)\right]
=\pm 2/(\pi z)$, and hence the vacuum stresses are uniform:
\begin{equation}
p=-\frac{1}{\pi }\int_{m}^{\infty }\frac{z^{2}dz}{\sqrt{z^{2}-m^{2}}}\frac{1%
}{e^{4az}\left( \frac{B_{1}z-A_{1}}{B_{1}z+A_{1}}\right) ^{2}-1}.
\label{pD2}
\end{equation}
For $B_{1}=0$ and $A_{1}=0$ we obtain from here the standard results for
Dirichlet and Neumann scalars. Unlike the stress distribution 
the energy distribution
is nonuniform. The corresponding expression directly follows from 
Eq.(\ref{q1in}):
\begin{equation}
\varepsilon =-\frac{1}{\pi }\int_{m}^{\infty }dz\frac{\sqrt{z^{2}-m^{2}}}{%
e^{4az}\left( \frac{B_{1}z-A_1}{B_{1}z+A_1}\right) ^{2}-1}\left[
1+\left( 1+\frac{4\xi -1}{1-m^{2}/z^{2}}\right) \frac{B_{1}z-A_{1}}{%
B_{1}z+A_{1}}e^{2az}\cosh 2zr\right] .  \label{epsD2in}
\end{equation}
Note that for the case of massless field the expressions (\ref{pD2}) and (%
\ref{epsD2in}) coincide with the general formulas derived in
\cite{RomeoSah} for two - plate geometry if one takes there $\beta
_{1}=\beta _{2}=B_{1}/A_{1}$ and $D=1$. In $D=1$ our
assumption that all zeros for $\bar{J}_{\nu }(z)$ are real
corresponds to $B_1/A_1<0$.

\section{Total Casimir energy and vacuum forces inside a sphere}

\label{sec:inen}

In the previous section we considered the scalar vacuum densities and
stresses inside a sphere. Here we will concentrate on
the corresponding global
quantities. First of all note that by using formulae 
(\ref{fieldmodesum1}) and (\ref{vevEMT}) the unregularized 
00-component for the vacuum EMT may be
presented in the form
\begin{equation}
\langle 0|T_{00}(x)|0\rangle
=\frac{1}{2a^{3}S_{D}r^{n}}\sum_{l=0}^{\infty
}D_{l}\sum_{k=1}^{\infty }\frac{\lambda _{\nu ,k}^{3}T_{\nu
}(\lambda _{\nu ,k})}{\sqrt{\lambda _{\nu
,k}^{2}+m^{2}a^{2}}}f_{\nu }^{(0)}\left[ J_{\nu }(\lambda _{\nu
,k}r/a)\right] ,  \label{T00unreg}
\end{equation}
where we have introduced the notation
\begin{equation}
f_{\nu }^{(0)}\left[ f(y)\right] =(1-4\xi )\left[ f^{^{\prime }2}(y)-\frac{%
n}{y}f(y)f^{\prime }(y)+\left( \frac{\nu ^{2}}{y^{2}}-\frac{4\xi +1}{%
4\xi -1}\right) f^{2}(y)\right] +2\left( \frac{mr}{y}\right) ^{2}f^{2}(y).
\label{fnu0}
\end{equation}
By using the standard integrals involving the Bessel functions for the total
volume energy inside a spherical shell one finds
\begin{equation}
E_{{\rm in}}^{({\rm vol})}=\frac{1}{2a}\sum_{l=0}^{\infty
}D_{l}\sum_{k=1}^{\infty }\sqrt{\lambda _{\nu ,k}^{2}+m^{2}a^2}\left[ 1+%
\frac{(4\xi -1)A_{1}B\lambda _{\nu ,k}^{2}}{(\lambda _{\nu ,k}^{2}+
m^{2}a^2)[A^{2}+B^{2}(\lambda _{\nu ,k}^{2}-\nu ^{2})]}\right] .  \label{volinen}
\end{equation}
As we see in the general Robin case this energy differs from the
total vacuum energy inside a sphere:
\begin{equation}
E_{{\rm in}}=\frac{1}{2}\sum_{l=0}^{\infty }D_{l}\sum_{k=1}^{\infty }\omega
_{\nu ,k},\quad \omega _{\nu ,k}=\sqrt{\lambda _{\nu ,k}^{2}/a^2+m
^{2}}.  \label{totinen}
\end{equation}
We will interpret this difference as a result of the presence of 
an additional surface energy contribution:
\begin{equation}
E_{{\rm in}}^{({\rm surf})}=-\frac{(4\xi
-1)A_{1}B}{2a}\sum_{l=0}^{\infty }D_{l}\sum_{k=1}^{\infty
}\frac{\lambda _{\nu ,k}^{2}}{\sqrt{\lambda _{\nu
,k}^{2}+m^{2}a^2}[A^{2}+B^{2}(\lambda _{\nu ,k}^{2}-\nu ^{2})]} ,
\label{totsurfen}
\end{equation}
such that
\begin{equation}
E_{{\rm in}}=E_{{\rm in}}^{({\rm vol})}+E_{{\rm in}}^{({\rm surf})}.
\label{voltotsurin}
\end{equation}
The surface energy (\ref{totsurfen}) can be obtained independently by
integrating the corresponding surface energy density. To see this let us note
that there is a surface energy density contribution to the total energy
density in the form \cite{Kennedy}
\begin{equation}
T_{00}^{({\rm surf})}=-(2\xi -1/2)\delta (x,\partial M)\varphi
n^{i}\partial _{i}\varphi ,  \label{surfendens}
\end{equation}
where $\delta (x,\partial M)$ - is a ''one sided'' $\delta $-distribution.
By using Eq.(\ref{fieldmodesum1}) the corresponding VEV takes the form
\begin{eqnarray}
\langle 0|T_{00}^{({\rm surf})}(x)|0\rangle &=&(2\xi -1/2)\delta
(x,\partial M)\left( \partial _{r}\langle 0|\varphi (r)\varphi (r^{\prime
})|0\rangle \right) _{r^{\prime }=r}=-(4\xi -1)\frac{\delta (r-a+0)}{%
2aS_{D}r^{D-1}}  \nonumber \\
&&\times \sum_{l=0}^{\infty }D_{l}\sum_{k=1}^{\infty
}\frac{\lambda _{\nu ,k}T_{\nu }(\lambda _{\nu
,k})}{\sqrt{\lambda _{\nu ,k}^{2}+m^{2}a^2}}J_{\nu }(z)\left[
\frac{n}{2}J_{\nu }(z)-zJ_{\nu }^{\prime }(z)\right] ,\quad
z=\lambda _{\nu ,k}r/a .  \label{surfendens1}
\end{eqnarray}
Integrating this density over the region inside a sphere, we obtain the
corresponding surface energy:
\begin{equation}
E_{{\rm in}}^{({\rm surf})}=\int d^{D}x\,\langle 0|T_{00}^{({\rm surf}%
)}(x)|0\rangle .  \label{totsurfen1}
\end{equation}
It can easily be seen that this expression with Eq.(\ref{surfendens1})
coincides with Eq.(\ref{totsurfen}).

The subtracted surface energy density can be obtained by applying to the sum
over $k$ in Eq.(\ref{surfendens1}) the 
summation formula (\ref{sumJ1anal}) and
omitting the term coming from the first integral on the right of this
formula. An alternative way is to use in Eq.(\ref{surfendens1}) 
the subtracted
product $\left\langle \varphi (r)\varphi (r^{\prime })\right\rangle _{SUB}$
from Eq.(\ref{regWight}) instead of $\langle 0|\varphi (r)\varphi (r^{\prime
})|0\rangle $. As a result one obtains
\begin{equation}
\left\langle T_{00}^{({\rm surf})}(x)\right\rangle _{SUB}=\delta (r-a+0)%
\frac{1-4\xi }{2\pi aS_{D}r^{D-1}}\sum_{l=0}^{\infty
}D_{l}\int_{ma
}^{\infty }\frac{zdz}{\sqrt{z^{2}-m^{2}a^2}}\frac{\bar{K}_{\nu }(z)}{\bar{I}%
_{\nu }(z)}F_{{\rm s}}[I_{\nu }(zr/a)],
\end{equation}
where for a given function $f(y)$ we have introduced thenotation
\begin{equation}
F_{{\rm s}}[f(y)]=f(y)\left[ yf^{\prime }(y)-\frac{n}{2}f(y)\right] .
\label{Fs}
\end{equation}
Integration of this energy density gives the total subtracted
surface energy:
\begin{equation}
E_{{\rm in}}^{({\rm surf})}=\frac{1-4\xi
}{2\pi a}\sum_{l=0}^{\infty }D_{l}\int_{ma }^{\infty }\frac{zdz}{\sqrt{%
z^{2}-m^{2}a^2}}\frac{\bar{K}_{\nu }(z)}{\bar{I}_{\nu }(z)}F_{{\rm s}%
}[I_{\nu }(z)] .  \label{surfsuben}
\end{equation}

Integrating the energy density $q=\varepsilon $ in Eq.(\ref{q1in}) over the
volume inside a sphere we obtain the corresponding subtracted volume energy.
Using the result that for any modified cylinder function $Z_{\nu }(y)=\
c_{1}I_{\nu }(y)+c_{2}K_{\nu }(y)$ we have the formula
\begin{equation}
\int dr\,rF_{{\rm \nu }}^{(\varepsilon )}[Z_{\nu }(zr)]=\frac{1}{z^{2}}F_{%
{\rm v}}[Z_{\nu }(zr)]  \label{enintform}
\end{equation}
with
\begin{equation}
F_{{\rm v}}[f(y)]=(1-4\xi )F_{{\rm s}}[f(y)]+\left( y^2-m^{2}a^2
\right) \left[ f^{^{\prime }2}(y)-\left( 1+\frac{\nu ^{2}}{y^{2}}%
\right) f^{2}(y)\right] , \label{Fvint1}
\end{equation}
one obtains
\begin{equation}
E_{{\rm in}}^{({\rm vol})}=-\frac{1}{2\pi a}%
\sum_{l=0}^{\infty }D_{l}\int_{ma }^{\infty }\frac{\bar{K}_{\nu }(z)}{\bar{I%
}_{\nu }(z)}\frac{zdz}{\sqrt{z^{2}-m^{2}a^2}}F_{{\rm v}}[I_{\nu }(z)] .
\label{totinenergy}
\end{equation}
Now, using relations (\ref{surfsuben}) and (\ref {totinenergy})
for the total Casimir energy inside a sphere, we get
\begin{equation}
E_{{\rm in}}=-\frac{1}{2\pi a}%
\sum_{l=0}^{\infty }D_{l}\int_{ma}^{\infty }\frac{dz}{z}\,\sqrt{%
z^{2}-m^{2}a^{2}}\frac{\bar{K}_{\nu }(z)}{\bar{I}_{\nu }(z)}
F_{{\rm t}}[I_{\nu }(z)],  \label{totsuben}
\end{equation}
where for future convenience we have defined
\begin{equation}
F_{{\rm t}}[f(y)]=y^{2}f^{^{\prime }2}(y)-\left( y^{2}+
\nu ^{2}\right) f_{\nu }^{2}(y).  \label{Ftn}
\end{equation}
As we see the dependences on the coupling parameter $\xi $ in
volume and surface energies canceled each other and the total
energy does not depend on this parameter. We might expect this
result as the eigenmodes do not depend on $\xi $. With the help
of the recurrence relations for $I_{\nu }(z)$ it can be seen that
for the case $D=3$ the Dirichlet massless scalar field interior
energy (\ref{totsuben}) coincides with the result given in
\cite{BenderHays}. The force acting per unit area of the sphere
from inside is determined as
\begin{equation}
F_{{\rm in}}=p|_{r=a-}=\frac{-1}{2\pi
a^{D+1}S_{D}}\sum_{l=0}^{\infty
}D_{l}\int_{ma }^{\infty }\frac{%
z^{3}dz}{\sqrt{z^{2}-m^{2}a^2}}\frac{\bar{K}_{\nu
}(z)}{\bar{I}_{\nu } (z)}F_{\nu }^{(p)}[I_{\nu }(z)]  ,
\label{Fin}
\end{equation}
with the notation (\ref{Finperad}). Note that one has the relation
\begin{equation}
z^2F_\nu ^{(p)}=F_{{\rm t}}[f(z)]+\xi _1F_{{\rm s}}[f(z)].
\label{Fpts}
\end{equation}
For the massless scalar this leads to the formula
\begin{equation}
a^{D}S_DF_{{\rm in}}=E_{{\rm in}}^{{\rm (vol)}}+\frac{4D}{4\xi
-1} (\xi -\xi _c)E_{{\rm in}}^{{\rm (surf)}} . \label{relFEin}
\end{equation}
This relation can also be derived directly from continuity
equation (\ref{conteq}), taking into account the formula
\begin{equation}
(D-1)p_{\perp }=\varepsilon -p-D(\xi -\xi _c)\nabla _i\nabla ^i
\langle \varphi ^2 \rangle _{{\rm reg}} , \label{pperbtrace}
\end{equation}
where we have used expression (\ref{trace}) for the EMT trace in
the case of a massless field. Substituting Eq.(\ref{pperbtrace}) into
Eq.(\ref{conteq}) and integrating over the interior region we obtain
Eq.(\ref{relFEin}). Note that for the minimally coupled scalar ($\xi
=0$) this formula may be presented in the form of the standard
thermodynamic relation
\begin{equation}
dE_{{\rm in}}^{{\rm (vol)}}=-F_{{\rm in}}dv_{D}+\sigma _{{\rm
in}} ds_{D}, \label{thermrelin}
\end{equation}
where $v_{D}=S_Da^{D}/D$ and $s_{D}=S_D a^{D-1}$ are the volume
and surface area for a $D$-dimensional ball with radius $a$, and
$\sigma _{{\rm in}}=E_{{\rm in}}^{{\rm (surf)}}/s_{D}$ is the
inner surface tension.

The expressions (\ref{surfsuben}), (\ref{totinenergy}), (\ref{totsuben})
and (\ref{Fin}) are divergent in the given forms. Moreover, in the sums
over $l$ the individual integrals diverge. To extract finite results we
can apply the procedure already used in \cite{MiltonSc,MiltonVec}. A
detailed description of this and the results for the corresponding
numerical evaluations will be reported elsewhere. Here we just briefly
outline the scheme with the example of the surface energy (\ref{surfsuben}).

First of all we note that \cite{MiltonSc}
\begin{equation}
\sum_{l=0}^{\infty }D_l=0,\quad D<1.
\label{sumDl}
\end{equation}
As a result for these values $D$ we can add to the integrand any
$l$-independent term without changing the value of the sum. We
choose this term in a way that makes the $z$-integral finite. For this we
write the asymptotic behavior of the integrand in
Eq.(\ref{surfsuben}) as $z\to \infty $:
\begin{equation}
\frac{\bar{K}_{\nu }(z)}{\bar{I}_{\nu }(z)}F_{\mathrm{s}}[I_{\nu }(z)]\sim
\left( \delta _{B0}-\frac{1}{2}\right) \left( 1-\frac{S^{(as)}}{2z}+\cdots
\right) ,  \label{surfintas}
\end{equation}
where we introduced the notation
\begin{equation}
S^{(as)}=\left\{
\begin{array}{cc}
D-3+4A/B, & B\neq 0 \\
D-1, & B=0 .
\end{array}
\right. .  \label{SurfSas}
\end{equation}
Note that the displayed terms are independent of $l$. Hence,
without changing the value of the surface energy for $D<1$ we can
write expression (\ref{surfsuben}) as
\begin{equation}
E_{\mathrm{in}}^{(\mathrm{surf})}=\frac{1-4\zeta }{2\pi a}\sum_{l=0}^{\infty
}D_{l}\int_{ma }^{\infty }\frac{zdz}{\sqrt{z^{2}-m^{2}a^2}}\left\{ \frac{%
\bar{K}_{\nu }(z)}{\bar{I}_{\nu }(z)}F_{\mathrm{s}}[I_{\nu }(z)]-\left(
\delta _{B0}-\frac{1}{2}\right) \left( 1-\frac{S^{(as)}}{2z}\right) \right\}
.  \label{Einsurfnew}
\end{equation}
In this form the integrals are well defined for a fixed $l$, as
for large $z\to \infty $ the subintegrand behaves as $z^{-2}$.
Then we analytically continue expression (\ref{Einsurfnew}) to
all $D$. To turn the sum over $l$ into a convergent series we can
use procedure often used in the previous literature on the
Casimir effect (see, for instance,
\cite{RomPR,Leseduarte,MiltonSc,MiltonVec,BorEK}). First we
rescale the integration variable, $z\rightarrow z\nu $. Then we
add and subtract the leading terms in the asymptotic expansion of
the subintegrand for $l\rightarrow \infty $ using the uniform
asymptotic expansions for the modified Bessel functions. The
subtracted part is finite and may be numerically evaluated, and
the asymptotic part can be expressed via Riemann or Hurwitz zeta
functions. Note that in the final result divergent terms in
the form of poles for the latter functions may remain. Here one
should take into account that the local cutoff dependent parts are
automatically lost in this regularization procedure. The
corresponding terms may be important in comparing the results
with experiments (for instance, in the electromagnetic case, see,
e.g., \cite{Candelas,Marach}), and can be extracted by expanding
the local expectation values (\ref{q1in}) in powers of the distance
from the boundary.

\section{Vacuum outside a sphere}
\label{sec:outdens}

To obtain the VEV for the EMT outside a sphere we consider first the
scalar vacuum in the layer between two concentric
spheres with radii $a$ and $b$, $a<b$%
. The corresponding boundary conditions have the form
\begin{equation}
\left( \tilde{A}_{r}+\tilde{B}_{r}\frac{\partial }{\partial
r}\right) \varphi (x) =0,\quad r=a,b,  \label{twospherebc}
\end{equation}
with constant coefficients $\tilde{A}_{r}$ and $\tilde{B}_{r}$,
in general, different for the inner and outer spheres. Now the
complete set of solutions to the field equation has the form
(\ref {eigfunc}) with the replacement
\begin{equation}
J_{\nu }(\lambda r)\rightarrow g_{\nu }(\lambda a,\lambda r)\equiv J_{\nu
}(\lambda r)\bar{Y}_{\nu }^{(a)}(\lambda a)-\bar{J}_{\nu }^{(a)}(\lambda
a)Y_{\nu }(\lambda r),\quad \lambda =\sqrt{\omega ^{2}-m^{2}},  \label{genu}
\end{equation}
where $Y_{\nu }(z)$ is the Neumann function,
and functions with overbars are defined in
analogy to Eq.(\ref{barnot}):
\begin{equation}
\bar{F}^{(\alpha )}(z)\equiv A_{\alpha }F(z)+B_{\alpha }zF^{\prime
}(z),\quad A_{\alpha }=\tilde{A}_{\alpha }-\tilde{B}_{\alpha }n/2\alpha
,\quad B_{\alpha }=\tilde{B}_{\alpha }/\alpha ,\;\alpha =a,b.
\label{barnotab}
\end{equation}
The eigenfunctions (\ref{genu}) satisfy the boundary condition 
on the sphere $r=a$.
From the boundary condition on $r=b$ one obtains that the corresponding
eigenmodes are solutions to the equation
\begin{equation}
C_{\nu }^{ab}(b/a,\lambda a)\equiv \bar{J}_{\nu }^{(a)}(\lambda a)\bar{Y}%
_{\nu }^{(b)}(\lambda b)-\bar{J}_{\nu }^{(b)}(\lambda b)\bar{Y}_{\nu
}^{(a)}(\lambda a)=0.  \label{eigmodesab}
\end{equation}
The coefficients $\beta _{\alpha }$ are determined from the normalization
condition (\ref{normcoef}), where now the integration goes over the region
between the spheres, $a\leq r\leq b$. Using the formula
for integrals involving
the product of any two cylinder functions one obtains
\begin{equation}
\beta _{\alpha }^{2}=\frac{\pi ^{2}\lambda }{4N(m_{k})\omega a}T_{\nu
}^{ab}(b/a ,\lambda a),  \label{betanormab}
\end{equation}
where $N(m_{k})$ comes from the normalization integral
(\ref{harmint}) and we use notation (\ref{tekaAB}).

Substituting the eigenfunctions into the mode sum
(\ref{fieldmodesum}) and using addition formula (\ref{adtheorem})
for the expectation value of the field product one finds
\begin{equation}
\langle 0|\varphi (x)\varphi (x^{\prime })|0\rangle =\frac{\pi
^{2}(rr^{\prime })^{-n/2}}{4naS_{D}}\sum_{l=0}^{\infty
}(2l+n)C_{l}^{n/2}(\cos \theta ) \sum_{k=1}^{\infty }h(\gamma
_{\nu ,k})T_{\nu }^{ab}(b/a, \gamma _{\nu ,k}),
\label{fieldmodesum1ab}
\end{equation}
with $\gamma _{\nu ,k} =\lambda a$ being the solutions to
Eq.(\ref{eigmodesab}) (see also the Appendix) and
\begin{equation}
h(z)=\frac{z}{\sqrt{z^{2}+m^{2}a^{2}}}g_{\nu }(z,zr/a)g_{\nu
}(z,zr^{\prime }/a)e^{i\sqrt{z^{2}/a^{2}+m^{2}}(t^{\prime }-t)}.
\label{hab}
\end{equation}
To sum over $k$ we will use the summation formula (\ref{cor3form}).
The corresponding conditions are satisfied if $%
r+r^{\prime }+|t-t^{\prime }|<2b$. Note that this is the case in
the coincidence limit for the region under consideration. Applying to the
sum over $k$ in Eq.(\ref{fieldmodesum1ab}) formula (\ref{cor3form}) one
obtains
\begin{equation}
\langle 0|\varphi (x)\varphi (x^{\prime })|0\rangle =\frac{1}{2naS_{D}}%
\sum_{l=0}^{\infty }\frac{2l+n}{(rr^{\prime })^{n/2}}C_{l}^{n/2}(\cos \theta
)\left\{ \int_{0}^{\infty }\frac{h(z)dz}{\bar{J}_{\nu }^{(a)2}(z)+\bar{Y}%
_{\nu }^{(a)2}(z)}\right.   \label{unregWightab}
\end{equation}
\[
-\left. \frac{2}{\pi }\int_{ma }^{\infty }\frac{zdz}{\sqrt{z^{2}-a^2m^{2}}}%
\frac{\bar{K}_{\nu }^{(b)}(\eta z)}{\bar{K}_{\nu }^{(a)}(z)}\frac{G_{\nu
}^{(a)}(z,zr/a)G_{\nu }^{(a)}(z,zr^{\prime }/a)}{\bar{K}_{\nu }^{(a)}(z)\bar{%
I}_{\nu }^{(b)}(\eta z)-\bar{K}_{\nu }^{(b)}(\eta z)\bar{I}_{\nu }^{(a)}(z)}%
\cosh \left[ \sqrt{z^{2}/a^{2}-m^{2}}(t^{\prime }-t)\right] \right\} ,
\]
where we have introduced notations
\begin{equation}
G_{\nu }^{(\alpha )}(z,y)=I_{\nu }(y)\bar{K}_{\nu }^{(\alpha )}(z)-\bar{I}%
_{\nu }^{(\alpha )}(z)K_{\nu }(y),\;\alpha =a,b  \label{Geab}
\end{equation}
(the function with $\alpha =b$ will be used below) with the
modified Bessel functions. Note that we have assumed values
$A_{\alpha }$ and $B_{\alpha }$ for which all zeros for
Eq.(\ref{eigmodesab}) are real and have omitted the residue terms
(\ref{comppoles}). In the following we will consider this case
only.

To obtain the vacuum EMT components outside a single sphere let us consider
the limit $b\rightarrow \infty $. In this limit the second integral on the
right of Eq.(\ref{unregWightab}) tends to zero (for large $b/a$ the
subintegrand is proportional to $e^{-2bz/a}$), whereas the first one does
not depend on $b$. It follows from here that the quantity
\begin{eqnarray}
\langle 0|\varphi (x)\varphi (x^{\prime })|0\rangle &=&\frac{1}{2naS_{D}}%
\sum_{l=0}^{\infty }\frac{2l+n}{(rr^{\prime })^{n/2}}C_{l}^{n/2}(\cos \theta
)  \label{unregWightout} \\
&\times &\int_{0}^{\infty
}\frac{zdz}{\sqrt{z^{2}+m^{2}a^{2}}}{\frac{g_{\nu }(z,zr/a)g_{\nu
}(z,zr^{\prime }/a)}{\bar{J}_{\nu }^{(a)2}(z)+\bar{Y}_{\nu
}^{(a)2}(z)}}e^{i\sqrt{z^{2}/a^{2}+m^{2}}(t^{\prime }-t)}
\nonumber
\end{eqnarray}
is the Wightman function for the exterior region of a single sphere with
radius $a$. To regularize this expression we have to subtract the
corresponding part for the unbounded space, which, as we
saw, can be presented in the form (\ref{Mink}). Using the relation
\begin{equation}
\frac{g_{\nu }(z,zr/a)g_{\nu }(z,zr^{\prime }/a)}{\bar{J}_{\nu }^{(a)2}(z)+%
\bar{Y}_{\nu }^{(a)2}(z)}-J_{\nu }(zr/a)J_{\nu }(zr^{\prime }/a)=-\frac{1}{2}%
\sum_{\sigma =1}^{2}\frac{\bar{J}_{\nu }^{(a)}(z)}{\bar{H}_{\nu }^{(\sigma
a)}(z)}H_{\nu }^{(\sigma )}(zr/a)H_{\nu }^{(\sigma )}(zr^{\prime }/a)
\label{relab}
\end{equation}
with $H_{\nu }^{(\sigma )}(z)$, $\sigma =1,2$, being the Hankel functions,
one obtains
\begin{eqnarray}
&&\langle 0|\varphi (x)\varphi (x^{\prime })|0\rangle -\langle
0_{M}|\varphi (x)\varphi (x^{\prime })|0_{M}\rangle
=-\frac{1}{4nS_{D}}\sum_{l=0}^{\infty }\frac{2l+n}{(rr^{\prime
})^{n/2}}C_{l}^{n/2}(\cos \theta )
\label{extdif} \\
&& \times \sum_{\sigma =1}^{2}\int_{0}^{\infty }{dz\,z\frac{e^{i\sqrt{%
z^{2}+m^{2}}(t^{\prime }-t)}}{\sqrt{z^{2}+m^{2}}}\frac{\bar{J}_{\nu
}^{(a)}(za)}{\bar{H}_{\nu }^{(\sigma a)}(za)}H_{\nu }^{(
\sigma )}(zr)H_{\nu }^{(\sigma )}(zr^{\prime }).}  \nonumber
\end{eqnarray}
Assuming that the function $\bar H_{\nu }^{(1a)}(z)$, 
($\bar H_{\nu }^{(2a)}(z)$)
has no zeros for $0<{\rm arg}\,z\leq \pi /2$ ($-\pi /2\leq {\rm arg}\,z<0$)
we can rotate the integration contour on the
right by the angle $\pi /2$ for $%
\sigma =1$ and by the angle $-\pi /2$ for $\sigma =2$. The integrals over $%
(0,ima )$ and $(0,-ima )$ cancel out and after introducing the
Bessel modified functions one obtains
\begin{eqnarray}
\langle \varphi (x)\varphi (x^{\prime })\rangle _{{\rm reg}} &=&-\frac{1}{%
n\pi S_{D}}\sum_{l=0}^{\infty }\frac{2l+n}{(rr^{\prime })^{n/2}}%
C_{l}^{n/2}(\cos \theta )  \label{regWightout} \\
&\times &\int_{m}^{\infty }{dz\,\ z\frac{\bar{I}_{\nu }^{(a)}(za)}{\bar{K}%
_{\nu }^{(a)}(za)}\frac{K_{\nu }(zr)K_{\nu }(zr^{\prime })}{\sqrt{z^{2}-m^{2}%
}}\cosh \!}\left[ {\sqrt{z^{2}-m^{2}}(t^{\prime }-t)}\right] .
\nonumber
\end{eqnarray}
For the VEV of the field square in the outside region this leads to
\begin{equation}
\langle \varphi ^{2}(r)\rangle _{{\rm reg}}=-\frac{1}{\pi ar^{n}S_{D}}%
\sum_{l=0}^{\infty }D_{l}\int_{ma}^{\infty }dz\ z\frac{\bar{I}_{\nu
}^{(a)}(z)}{\bar{K}_{\nu }^{(a)}(z)}\frac{K_{\nu }^{2}(zr/a)}{\sqrt{%
z^{2}-m^{2}a^{2}}},  \label{regsquareout}
\end{equation}
where $D_{l}$ is defined in accord with Eq.(\ref{Dlang}).

As in the interior case the vacuum EMT is diagonal and the
corresponding components can be presented in the form
\begin{equation}
q(a,r)=-\frac{1}{2\pi a^{3}r^{n}S_{D}}\sum_{l=0}^{\infty
}D_{l}\int_{ma}^{\infty }dz\,z^{3}\frac{\bar{I}_{\nu }^{(a)}(z)}{\bar{K}%
_{\nu }^{(a)}(z)}\frac{F_{\nu }^{(q)}\left[ K_{\nu }(zr/a)\right] }{\sqrt{%
z^{2}-m^{2}a^{2}}},\quad q=\varepsilon ,\,p,\,p_{\perp },\quad r>a,
\label{q1out}
\end{equation}
where the functions $F_{\nu }^{(q)}\left[ f(y)\right] $ are given
by relations (\ref{Fineps}), (\ref{Finperad}) and (\ref{Finpeaz}).
As for the interior components, the quantities (\ref{q1out})
diverge at the sphere surface, $r=a$. The leading terms of the
asymptotic expansions are determined by same formulas
(\ref{peinasimp}) and (\ref{epsinasimp}) with the replacement
$a-r\rightarrow r-a$.

When $D=1$ from Eq.(\ref{q1out}) we obtain the expectation values
for the one dimensional semi-infinite region $r>a$. Now the
consideration similar to (\ref{pD2}) and (\ref{epsD2in}) yields
\begin{equation}\label{epsD2out}
  \varepsilon =-\frac{1}{2\pi }\int_{m}^{\infty }dz
  \frac{4\xi z^{2}-m^{2}}{\sqrt{z^{2}-m^{2}}}
  \frac{\tilde{B}_{a}z+\tilde{A}_{a}}{\tilde{B}_{a}z-\tilde{A}_{a}}
  e^{-2z(r-a)},\quad p=0,\quad D=1.
\end{equation}
These results can also be obtained from Eq.(\ref{epsD2in}) and
(\ref{pD2}) in the limit $a\to \infty $, $r+a={{\rm const}}$ (in
these formulas $r+a$ is the distance from the boundary and
corresponds to $r-a$ in Eq.(\ref{epsD2out})). For the massless
scalar Eq.(\ref{epsD2out}) is a special case of the general
formula from \cite{RomeoSah} for a Robin single plate in an
arbitrary dimension.

In the case of a massless scalar the asymptotic behavior for
expectation values (\ref{q1out}) at large distances from the
sphere can be obtained by introducing a new integration variable
$y=zr/a$ and expanding the subintegrand in terms of $a/r$. The
leading contribution for the summand with a given $l$ has 
order $(a/r)^{2l+2D-1}$ (assuming that $A_{a}\neq \nu B_{a}$) and
the main contribution comes from the $l=0$ term. As a result we can
see that for $r\gg a$
\begin{equation}\label{relfar1}
  \varepsilon \approx p\approx -\frac{D-2}{D-1}p_{\perp },
\end{equation}
with
\begin{equation}\label{relfarD2}
  \varepsilon \approx \frac{\xi -\xi _{c}}{8r^{3}\ln r/a},\quad
  D=2 ,
\end{equation}
and
\begin{equation}\label{relfarD}
  \varepsilon \approx (\xi -\xi _{c})
  \frac{\tilde{A}_{a}(D-2)}{\tilde{A}_{a}-B_{a}n}
  \frac{\Gamma (D-1/2)\Gamma \left( \frac{D+1}{2}\right) }{2^{D-1}
  \pi ^{D/2}\Gamma ^{2}(D/2)}\frac{a^{D-2}}{r^{2D-1}},\quad D>2.
\end{equation}
The case $D=1$ simply follows from Eq.(\ref{epsD2out}). For the
conformally coupled scalar the vacuum densities behave as
$1/r^{2D+1}$.

In analogy to the interior case the outside subtracted surface
energy density can be presented in the form
\begin{equation}
\left\langle T_{00}^{({\rm surf})}(x)\right\rangle _{SUB}=\delta (r-a-0)%
\frac{4\xi -1}{2\pi aS_{D}r^{D-1}}\sum_{l=0}^{\infty
}D_{l}\int_{ma}^{\infty }\frac{zdz}{\sqrt{z^{2}-m^{2}a^{2}}}\frac{\bar{I}%
_{\nu }^{(a)}(z)}{\bar{K}_{\nu }^{(a)}(z)}F_{{\rm s}}[K_{\nu }(zr/a)],
\label{surfsubdensout}
\end{equation}
where $F_{{\rm s}}[f(y)]$ is defined as in Eq.(\ref{Fs}). 
Integration of this formula gives the total surface energy
\begin{equation}
E_{{\rm ext}}^{({\rm surf})}=\frac{4\xi -1%
}{2\pi a}\sum_{l=0}^{\infty }D_{l}\int_{ma}^{\infty }\frac{zdz}{\sqrt{%
z^{2}-m^{2}a^{2}}}\frac{\bar{I}_{\nu }^{(a)}(z)}{\bar{K}_{\nu }^{(a)}(z)}
F_{{\rm s}}[K_{\nu }(z)],  \label{surfsubenout}
\end{equation}
localized on the outer surface of the sphere.

Integrating the energy density (\ref{q1out}), $q=\varepsilon $,
over the volume outside a sphere we obtain the corresponding
volume energy:
\begin{equation}
E_{{\rm ext}}^{({\rm vol})}=\frac{1}{2\pi a}%
\sum_{l=0}^{\infty }D_{l}\int_{ma}^{\infty }\frac{zdz}{\sqrt{z^{2}-m^{2}a^{2}%
}}\frac{\bar{I}_{\nu }^{(a)}(z)}{\bar{K}_{\nu }^{(a)}(z)}F_{{\rm
v}}[K_{\nu }(z)] , \label{totoutenergy}
\end{equation}
with the notation (\ref{Fvint1}). Now the 
total vacuum energy for the exterior
region is obtained as the sum of the volume and surface parts,
\begin{equation}
E_{{\rm ext}}=E_{{\rm ext}}^{({\rm vol})}+E_{{\rm ext}}^{({\rm surf}%
)}=
\frac{1}{2\pi a}\sum_{l=0}^{\infty }D_{l}\int_{ma}^{\infty }\frac{dz}{z}
\sqrt{z^{2}-m^{2}a^{2}}\frac{\bar{I}_{\nu }^{(a)}(z)}{\bar{K}_{\nu }^{(a)}
(z)} F_{{\rm t}}[K_{\nu }(z)],  \label{Etotout}
\end{equation}
where $F_{{\rm t}}[f(y)]$ is defined in accordance with
Eq.(\ref{Ftn}). As we see, as in the interior case, the
dependences on the curvature coupling $\xi $ in the surface and
volume energies cancel each other.

The radial projection of the force acting per unit area of the sphere from
outside is determined as
\begin{equation}
F_{{\rm ext}}=-p|_{r=a+}=\frac{1}{2\pi
a^{D+1}S_{D}}\sum_{l=0}^{\infty
}D_{l}\int_{ma}^{\infty }\frac{z^{3}dz}{\sqrt{z^{2}-m^{2}a^{2}}}\frac{\bar{I}%
_{\nu }^{(a)}(z)}{\bar{K}_{\nu }^{(a)}(z)}F_{\nu }^{(p)}[K_{\nu }(z)].
\label{Fout}
\end{equation}
Using relation (\ref{Fpts}), for the massless
scalar field one finds
\begin{equation}
a^{D}S_DF_{{\rm ext}}=E_{{\rm ext}}^{{\rm (vol)}}+\frac{4D}{4\xi
-1} (\xi -\xi _c)E_{{\rm ext}}^{{\rm (surf)}} . \label{relFEext}
\end{equation}
As in the interior case, this relation can also be obtained
directly from the continuity equation (\ref{conteq}). Recall that the
second summand on the right comes from the nonzero trace for the
nonconformally coupled massless scalar. The exterior surface,
volume, total energies, and vacuum forces acting on the sphere
written in the form (\ref{surfsubenout})-(\ref{Fout}) are
divergent. Here the corresponding scheme to extract finite
results is similar to that for the interior quantities and is
explained in section \ref{sec:inen}.

We now turn to the case of a \emph{spherical shell with zero thickness}.
The total vacuum
energy including the interior and exterior contributions can be 
obtained by summing Eqs.(\ref{totinenergy}) and
(\ref{totoutenergy}), and the resulting vacuum force by summing 
Eqs.(\ref{Fin}) and (\ref{Fout}):
\begin{equation}
E^{{\rm (vol)}}(a)=E_{{\rm in}}^{{\rm (vol)}}(a)+E_{{\rm ext}}^{{\rm (vol)}%
}(a),\quad E(a)=E_{{\rm in}}(a)+E_{{\rm ext}}(a),\quad F=F_{{\rm in}}+F_{%
{\rm ext}}.  \label{totenergy}
\end{equation}
Assuming that for the exterior and interior boundary conditions the
coefficients $A$ and $B$ are the same (this corresponds to $B_{1}=-B_{a}$ in
Eqs.(\ref{robcond}) and (\ref{twospherebc})) after some transformations the
corresponding expressions can be presented in the forms
\begin{equation}
E^{{\rm (surf)}}(a)=\frac{4\xi -1}{2\pi a}\sum_{l=0}^{\infty
}D_{l}\int_{ma}^{\infty }\frac{zdz}{\sqrt{z^{2}-m^{2}a^{2}}}\left[ 1-\left(
\beta +\frac{n}{2}\right) \frac{\left( \tilde{I}_{\nu }(z)\tilde{K}_{\nu
}(z)\right) ^{\prime }}{z\tilde{I}_{\nu }^{\prime }(z)\tilde{K}_{\nu
}^{\prime }(z)}\right] \;  \label{Etotsurf}
\end{equation}
for the total surface energy,
\begin{equation}
E(a)=-\frac{1}{2\pi a}\sum_{l=0}^{\infty }D_{l}\int_{ma}^{\infty }\frac{dz}{z%
}\sqrt{z^{2}-m^{2}a^{2}}\left[ 2\beta +\left( \nu ^{2}-\beta
^{2}+z^{2}\right) \frac{\left( \tilde{I}_{\nu }(z)\tilde{K}_{\nu }(z)\right)
^{\prime }}{z\tilde{I}_{\nu }^{\prime }(z)\tilde{K}_{\nu }^{\prime }(z)}%
\right]  \label{Etot}
\end{equation}
for the total energy, and
\begin{eqnarray}
F(a) &=& -\frac{1}{2\pi a^{D+1}S_{D}}
\sum_{l=0}^{\infty }D_{l}\int_{ma}^{\infty }%
\frac{zdz}{\sqrt{z^{2}-m^{2}a^{2}}}  \nonumber \\
&\times & \left[ 2\beta -\xi _{1}+\left[ \nu ^{2}-\beta
^{2}+z^{2}+\xi _{1}\left( \beta +\frac{n}{2}\right) \right] \frac{\left(
\tilde{I}_{\nu }(z)\tilde{K}_{\nu }(z)\right) ^{\prime
}}{z\tilde{I}_{\nu }^{\prime }(z)\tilde{K}_{\nu }^{\prime
}(z)}\right]  \label{Ftot}
\end{eqnarray}
for the resulting force per unit area acting on the spherical
shell. In these formulas we use the notations
\begin{equation}
\tilde{f}(z)=z^{\beta }f(z),\quad \beta =A/B,\quad  \label{ftilde}
\end{equation}
for a given function $f(z)$ and $\xi _1$ is defined in
Eq.(\ref{Finperad}). It can be easily seen that in the case $m=0$ and
$\beta =D/2-1$ formula (\ref{Etot}) coincides with the expression
for the Casimir energy of the TM modes derived in
\cite{MiltonVec}. The corresponding electromagnetic force can be
obtained by differentiating this expression with respect to $a$.

Using the relation
\begin{equation}
2\beta +(z^{2}+\nu ^{2}-\beta ^{2})\frac{\left( \tilde{I}_{\nu }(z)\tilde{K}%
_{\nu }(z)\right) ^{\prime }}{z\tilde{I}_{\nu }^{\prime }(z)\tilde{K}_{\nu
}^{\prime }(z)}=z\frac{\left( \bar{I}_{\nu }(z)\bar{K}_{\nu }(z)\right)
^{\prime }}{\bar{I}_{\nu }(z)\bar{K}_{\nu }(z)},  \label{reltildbar}
\end{equation}
the total Casimir energy (\ref{Etot}) can also be presented in the form
\begin{equation}
E(a)=-\frac{1}{2\pi a}\sum_{l=0}^{\infty }D_{l}\int_{ma}^{\infty }dz\sqrt{%
z^{2}-m^{2}a^{2}}\frac{\left( \bar{I}_{\nu }(z)\bar{K}_{\nu }(z)\right)
^{\prime }}{\bar{I}_{\nu }(z)\bar{K}_{\nu }(z)}.  \label{Etotbar}
\end{equation}
For the massless field, using the trace relation (\ref{pperbtrace}) 
and integrating the continuity equation (\ref{conteq}) one obtains
\begin{equation}
E^{{\rm vol}}=a^{D}S_D F-S_D \int _{0}^{\infty } dr\, r^{D-2}
\langle T_i^i\rangle _{{\rm reg}} . \label{Evoltrace}
\end{equation}
Note that though for the massless scalar the subintegrand is a
total divergence (see, Eq.(\ref{trace})) the corresponding contribution
to the right hand side of Eq.(\ref{Evoltrace}) is nonzero as the
function $(d/dr)\langle \varphi ^2\rangle _{{\rm reg}}$ is
discontinous at  $r=a$. This contribution can be expressed via
the total surface energy, like to Eq.(\ref{relFEin}) or
Eq.(\ref{relFEext}).

The formulas for the cases of Dirichlet and Neumann boundary
conditions can be obtained taking $B_{1}=0$ ($\beta =\infty $%
) and $A_{1}=0$ ($\beta =1-D/2$), respectively. For example, in the
case of Dirichlet vacuum force per unit area one has
\begin{equation}
F(a)=-\frac{1}{2\pi a^{D+1}S_{D}}\sum_{l=0}^{\infty }D_{l}\int_{ma}^{\infty }%
\frac{zdz}{\sqrt{z^{2}-m^{2}a^{2}}}\left[ 4(n+1)\xi -n+z
\frac{\left( I_{\nu
}(z)K_{\nu }(z)\right) ^{\prime }}{I_{\nu }(z)K_{\nu }(z)}\right] .
\label{FtotDirichlet}
\end{equation}
By using the expressions for $D_{l}$ and $S_{D}$ it can be seen that
in the case of a minimally coupled massless scalar ($\xi =0,\,m=0$)
the expression (\ref{FtotDirichlet}) coincides with the formula
derived in \cite{MiltonSc}.

For the Dirichlet and Neumann boundary conditions from
Eq.(\ref{Etotsurf}) one has
\begin{equation}
E_{(D)}^{{\rm (surf)}}=-E_{(N)}^{{\rm (surf)}}=
\frac{1-4\xi }{2\pi }\sum _{l=0}^{\infty }D_l\int _{m}^{\infty }
\frac{zdz}{\sqrt{z^2-m^2}}. \label{EtotsurfDN}
\end{equation}
These quantities are independent of the sphere radius and, if we
follow the regularization procedure developed in
\cite{MiltonSc,MiltonVec}, we conclude that for a thin
spherical shell the regularized surface energy vanishes for
Dirichlet and Neumann scalars. More generally, for Dirichlet and
Neumann boundary conditions the regularized interior,
Eq.(\ref{surfsuben}), and exterior, Eq.(\ref{surfsubenout}), surface
energies vanish separately. To see this note that VEV $\langle
0\vert \varphi (x)\partial _r'\varphi (x')\vert 0\rangle  $ is
zero for $r=a$ in the Dirichlet case and for $r'=a$ in the
Neumann case. As a result for these values the subtracted product
$\langle \varphi (x)\partial _r'\varphi (x')\rangle _{{\rm reg}}$
coincides with the corresponding Minkowskian part $\langle
0_M\vert \varphi (x)\partial _r'\varphi (x')\vert 0_M\rangle  $
(this can also be seen by direct evaluation, applying to the sum
over $l$ the Gegenbauer addition theorem, and taking the value
of the remained integral from \cite{Prudnikov}) and is zero in the
limit $r'\to a$ or $r\to a$ if we regularize the Minkowskian part
to be zero. Now we lead to our conclusion if we recall that the
surface energy is proportional to $\langle \varphi (a)\partial
_r\varphi (a)\rangle _{{\rm reg}}$. This is in agreement with the
regularization procedure from \cite{MiltonSc,MiltonVec}. First of
all, taking into account that the regularized value for the
integro-sum on the right of Eq.(\ref{EtotsurfDN}) is zero, we can
derive from Eqs.(\ref{surfsuben}) and (\ref{surfsubenout}) the
relations
\begin{equation}\label{EsurfinextDN}
  E_{{\rm in(D)}}^{{\rm (surf)}}=E_{{\rm in(N)}}^{{\rm (surf)}}
  =-E_{{\rm ext(D)}}^{{\rm (surf)}}=-E_{{\rm ext(N)}}^{{\rm
  (surf)}}.
\end{equation}
Hence, it is sufficient to consider the Dirichlet interior
energy. It can be written in the form
\begin{equation}\label{EsurfinD}
  \frac{E_{{\rm in(D)}}^{{\rm (surf)}}}{a^DS_D(4\xi -1)}=
  \lim _{r\to a-}\langle \varphi (a)\partial _r\varphi (r)
  \rangle _{{\rm reg}}=-\lim _{r\to a-}\partial _r
  \langle 0_M\vert \varphi (a)
  \varphi (r) \vert 0_M\rangle ,
\end{equation}
where we have used that for a Dirichlet scalar $\langle 0\vert
\varphi (a) \partial _r\varphi (r) \vert 0\rangle =0$. Now taking
into account expression (\ref{MinkWight}) for the Minkowskian
Wightmann function, expanding $K_{(D-1)/2}(m(a-r))$, and using
\begin{equation}\label{relfordimreg}
  (1-x)^{-\alpha }=\sum_{l=0}^{\infty }x^{l}
  \frac{\Gamma (l+\alpha )}{\Gamma (\alpha )l!},
\end{equation}
for the interior surface energy one obtains
\begin{equation}\label{EsurfinD1}
E_{{\rm in(D)}}^{{\rm (surf)}}=\frac{4\xi -1}{2^{D}a\Gamma (D/2)}
\sum_{k=0}^{[(D-1)/2]}\frac{(-1)^{k}(ma)^{2k}}{k!\Gamma (D/2-k)}
\sum_{l=0}^{\infty }\frac{\Gamma (l+D-2k)}{l!} ,
\end{equation}
where the square brackets denote the largest integer less than or
equal to its argument. By using the analytic continuation over
$D$ in \cite{MiltonSc} it has been shown that the regularized
value for the sum over $l$ in Eq.(\ref{EsurfinD1}) is zero, and
hence this is the case for the Dirichlet surface energy as well.

The method of extracting finite results from expressions
(\ref{Etotsurf})-(\ref{Ftot}) is the same as that for the
interior and exterior quantities and is explained in section
\ref{sec:inen}. In particular, subtracting the leading term in
the asymptotic expansion of the subintegrand for large $z$ and
using Eq.(\ref{sumDl}), the total Casimir energy may be presented in
the form
\begin{equation}
E=-\frac{1}{2\pi a}\sum_{l=0}^{\infty }D_{l}\int_{ma}^{\infty }dz\sqrt{%
z^{2}-m^{2}a^{2}}\frac{d}{dz}\ln \vert z^{2\delta _{B0}-1}
\bar{I}_{\nu }(z)\bar{K}_{\nu }(z)\vert .  \label{Etotbarnew}
\end{equation}
Here the individual integrals in the series are convergent. This
formula can be alternatively derived using, for instance, the
Green function method \cite{MiltonSc,MiltonVec} or applying to
the corresponding mode sum the methods already used in
\cite{RomPR}-\cite{BorEK},\cite{Nesterenko}-\cite{Cognola}. For
the case of a $D=3$ massless scalar with Neumann boundary condition
we recover the known result from \cite{Nesterenko}.

\section{Vacuum energy density and stresses in a spherical layer}
\label{sec:twosurfdens}

We have seen that in the intermediate region between two
concentric spheres the Wightman function for the scalar field can
be presented in the form (\ref {unregWightab}), where the term
with the first integral is the corresponding unregularized
function for the region outside a single sphere with radius $a$.
The regularization of this term was carried out in the previous
section. The second integral on the right of
Eq.(\ref{unregWightab}) will give a finite result at the coincidence limit
for $a\leq r<b$ and needs no regularization. As a result for the
regularized Wightman function in the region between two spheres
one obtains
\begin{eqnarray}
\langle \varphi (x)\varphi (x^{\prime })\rangle _{{\rm reg}} &=&\langle
\varphi (x)\varphi (x^{\prime })\rangle _{{\rm reg}}^{{\rm ext}}-\frac{1}{%
\pi nS_{D}}\sum_{l=0}^{\infty }\frac{2l+n}{(rr^{\prime })^{n/2}}%
C_{l}^{n/2}(\cos \theta )  \label{regWightab1} \\
&\times &\int_{m}^{\infty }\frac{zdz}{\sqrt{z^{2}-m^{2}}}\Omega _{a\nu
}(az,bz)G_{\nu }^{(a)}(az,zr)G_{\nu }^{(a)}(az,zr^{\prime })\cosh \left[
\sqrt{z^{2}-m^{2}}(t^{\prime }-t)\right] ,  \nonumber
\end{eqnarray}
where $\langle \varphi (x)\varphi (x^{\prime })\rangle _{{\rm reg}}^{{\rm ext%
}}$ is the corresponding function for the vacuum outside a single
sphere with radius $a$ given by Eq.(\ref{regWightout}), and we have
introduced the notation
\begin{equation}
\Omega _{a\nu }(az,bz)=\frac{\bar{K}_{\nu }^{(b)}(bz)/\bar{K}_{\nu
}^{(a)}(az)}{\bar{K}_{\nu }^{(a)}(az)\bar{I}_{\nu }^{(b)}(bz)-\bar{K}_{\nu
}^{(b)}(bz)\bar{I}_{\nu }^{(a)}(az)}.  \label{Omega}
\end{equation}
Substituting Eq.(\ref{regWightab1}) into Eq.(\ref{vevEMT}) 
we obtain that the
vacuum EMT has the diagonal form (\ref{diagEMT}) with components
\begin{equation}
q(a,b,r)=q(a,r)+q_{a}(a,b,r),\quad a<r<b,\quad q=\varepsilon
,p,p_{\perp },  \label{compab1}
\end{equation}
where $q(a,r)$ are the corresponding functions for the vacuum
outside a single sphere with radius $a$. In Eq.(\ref{compab1}) the
additional components are in the form
\begin{equation}
q_{a}(a,b,r)=-\frac{1}{2\pi r^{n}S_{D}}\sum_{l=0}^{\infty
}D_{l}\int_{m }^{\infty }\frac{z^{3}dz}{\sqrt{z^{2}-m^{2}}}\Omega
_{a\nu }(az,bz)F_{\nu }^{(q)}\left[ G_{\nu }^{(a)}(az,zr)\right]
,\quad q=\varepsilon ,\,p,\,p_{\perp },  \label{q1ab}
\end{equation}
where $F_{\nu }^{(q)}\left[ f(y)\right] $ is defined by relations
(\ref{Fineps}), (\ref{Finperad}), and (\ref{Finpeaz}). The
quantities (\ref{q1ab}) are finite for $a\leq r<b$ and diverge at
the surface $r=b$.

It can be seen that for the case of two spheres the Wightman function
in the intermediate region can also be
presented in the form
\begin{eqnarray}
\langle \varphi (x)\varphi (x^{\prime })\rangle _{{\rm reg}} &=&\langle
\varphi (x)\varphi (x^{\prime })\rangle _{{\rm reg}}^{{\rm in}}-\frac{1}{%
\pi nS_{D}}\sum_{l=0}^{\infty }\frac{2l+n}{(rr^{\prime })^{n/2}}%
C_{l}^{n/2}(\cos \theta )  \label{regWightab2} \\
&\times &\int_{m}^{\infty }\frac{zdz}{\sqrt{z^{2}-m^{2}}}\Omega _{b\nu
}(az,bz)G_{\nu }^{(b)}(bz,zr)G_{\nu }^{(b)}(bz,zr^{\prime })\cosh \left[
\sqrt{z^{2}-m^{2}}(t^{\prime }-t)\right] , \nonumber
\end{eqnarray}
with $\langle \varphi (x)\varphi (x^{\prime })\rangle _{{\rm reg}}^{{\rm int}%
}$ being the regularized Wightman function for the vacuum inside a single sphere
with radius $b$ (see Eq.(\ref{regWight}) with replacement $a\rightarrow b$),
and
\begin{equation}
\Omega _{b\nu }(az,bz)=\frac{\bar{I}_{\nu }^{(a)}(az)/\bar{I}_{\nu
}^{(b)}(bz)}{\bar{K}_{\nu }^{(a)}(az)\bar{I}_{\nu }^{(b)}(bz)-\bar{K}_{\nu
}^{(b)}(bz)\bar{I}_{\nu }^{(a)}(az)}.  \label{Omegatilde}
\end{equation}
Note that in the coincidence limit, $x'=x$, the second summand on
the right hand side of Eq.(\ref{regWightab1}) is finite on the
boundary $r=a$, and is divergent on the boundary $r=b$.
Similarly, the second summand on the right of formula
(\ref{regWightab2}) is finite on the boundary $r=b$ and diverges
on the boundary $r=a$. It follows from here that if we write the
regularized Wightmann function in the form
\begin{equation}
\langle \varphi (x)\varphi (x^{\prime })\rangle _{{\rm
reg}}=\langle \varphi (x)\varphi (x^{\prime })\rangle _{{\rm
reg}}^{{\rm in}}+\langle \varphi (x)\varphi (x^{\prime })\rangle
_{{\rm reg}}^{{\rm ext}}+\Delta W(x,x^{\prime }),
\label{intWeight}
\end{equation}
then in the coincidence limit the ''interference'' term $\Delta
W(x,x^{\prime })$ is finite for all values $a\leq r\leq b$. Using
formulas (\ref{regWight}) and (\ref{regWightab1}) it can be seen
that this term may be presented as
\begin{equation}
\Delta W(x,x^{\prime })=\frac{-1}{\pi nS_{D}}\sum_{l=0}^{\infty }\frac{2l+n}{%
(rr^{\prime })^{n/2}}C_{l}^{n/2}(\cos \theta )\int_{m}^{\infty }\frac{zdz}{%
\sqrt{z^{2}-m^{2}}}W^{(ab)}(r,r^{\prime })\cosh \left[ \sqrt{z^{2}-m^{2}}%
(t^{\prime }-t)\right] ,  \label{intWeightrr}
\end{equation}
where
\begin{equation}
W^{(ab)}(r,r^{\prime })=\frac{\bar{I}_{\nu }^{(a)}(az)}{\bar{I}_{\nu
}^{(b)}(bz)}\frac{\bar{K}_{\nu }^{(b)}(bz)}{\bar{K}_{\nu }^{(a)}(az)}\left[
\frac{G_{\nu }^{(a)}(az,zr)G_{\nu }^{(b)}(bz,zr^{\prime })}{\bar{K}_{\nu
}^{(a)}(az)\bar{I}_{\nu }^{(b)}(bz)-\bar{K}_{\nu }^{(b)}(bz)\bar{I}_{\nu
}^{(a)}(az)}-I_{\nu }(zr^{\prime })K_{\nu }(zr)\right] .  \label{Wrr}
\end{equation}

On the basis of formula (\ref{regWightab2}) the vacuum EMT
components may be written in another equivalent form:
\begin{equation}
q(a,b,r)=q(b,r)+q_{b}(a,b,r),\quad a<r<b,\quad q=\varepsilon
,p,p_{\perp },  \label{compab2}
\end{equation}
with $q(a,r)$ being the corresponding components for the vacuum
inside a single sphere with radius $b$ (expressions (\ref{q1in})
with replacement $a\rightarrow b$). Here the additional components
are given by the formula
\begin{equation}
q_{b}(a,b,r)=-\frac{1}{2\pi r^{n}S_{D}}\sum_{l=0}^{\infty
}D_{l}\int_{m}^{\infty }\frac{z^{3}dz}{\sqrt{z^{2}-m^{2}}}\Omega
_{b\nu }(az,bz)F_{\nu }^{(q)}\left[ G_{\nu }^{(b)}(bz,zr)\right]
,\quad q=\varepsilon ,\,p,\,p_{\perp }.  \label{q2ab}
\end{equation}
This expressions are finite for all $a<r\leq b$ and diverge at the inner sphere
surface $r=a$.

It follows from the above that if we present the vacuum EMT components in the
form
\begin{equation}
q(a,b,r)=q(a,r)+q(b,r)+\Delta q(a,b,r),\quad a<r<b,  \label{qinterf}
\end{equation}
then the quantities (no summation over $i$)
\begin{equation}
\Delta q(a,b,r)=q_{a}(a,b,r)-q(b,r)=q_{b}(a,b,r)-q(a,r)=
\hat{\theta }_{ii}\Delta W(x,x') \label{deltaq}
\end{equation}
are finite for all $a\leq r\leq b$, and $i=0,1,2$ correspond to
$q=\varepsilon , p,p_{\perp }$, respectively.
Near the surface $r=a$ it is suitable to
use the first equality in Eq.(\ref{deltaq}), as for $r\rightarrow a$ both
summands are finite. For the same reason the second equality is suitable for
calculations near the outer surface $r=b$. In particular, the additional radial
vacuum pressure on the sphere with $r=\alpha $, $\alpha =a,b$ due to the
existence of the second sphere (''interaction'' force ) can
be found from Eqs.(\ref{q1ab}) and (\ref{q2ab}), respectively. Using the
relations
\begin{equation}
G_{\nu }^{(r)}(rz,rz)=-B_{r},\quad rz\frac{\partial }{\partial y}
G_{\nu }^{(r)}(rz,y)\mid _{y=rz}=A_{r},\quad r=a,b,  \label{Grelab}
\end{equation}
they can be presented in the form
\begin{eqnarray}
p_{\alpha }(a,b,r=\alpha )&=&-\frac{1}{2\pi \alpha ^{D}S_{D}}%
\sum_{l=0}^{\infty }D_{l}\int_{m}^{\infty }\frac{zdz}{\sqrt{z^{2}-m^{2}}}%
\Omega _{\alpha \nu }(az,bz)   \label{paabb} \\
&\times & \left\{ \tilde{A}_{\alpha }^{2}-4(D-1)\xi
\tilde{A}_{\alpha } B_{\alpha }-\left[ z^{2}\alpha
^{2}+l(l+D-2)\right] B_{\alpha }^{2} \right\} ,\qquad \alpha =a,b.
\nonumber
\end{eqnarray}
Unlike the self-action forces this quantity is finite for $a<b$
and needs no further regularization. Using the Wronskian
\begin{equation}
\bar{K}_{\nu }^{(\alpha )}(\alpha z)\frac{d}{dz}\bar{I}_{\nu
}^{(\alpha )}(bz)-\bar{I}_{\nu }^{(\alpha )}(\alpha z)\frac{d}{dz}%
\bar{K}_{\nu }^{(\alpha )}(bz)=\frac{A_{\alpha }^2-B_{\alpha }^2(z^{2}\alpha
^{2}+\nu ^{2})}{z}  \label{wronskKI}
\end{equation}
it can be seen that
\begin{equation}
\left[ A_{\alpha }-B_{\alpha }(z^{2}\alpha ^{2}+\nu ^{2})\right] \Omega
_{\alpha \nu }(az,bz)=-n_{\alpha }\alpha \frac{\partial }{\partial \alpha }%
\ln \left| 1-\frac{\bar{I}_{\nu }^{(a)}(az)}{\bar{I}_{\nu }^{(b)}(bz)}\frac{%
\bar{K}_{\nu }^{(b)}(bz)}{\bar{K}_{\nu }^{(a)}(az)}\right| ,
  \label{Omeganew}
\end{equation}
where $n_{a}=1,\, n_{b}=-1$. This allows us to write the expressions
(\ref{paabb}) for the interaction forces per unit surface in
another equivalent form:
\begin{eqnarray}
p_{\alpha }(a,b,r=\alpha )&=&\frac{n_{\alpha }}{2\pi \alpha ^{D-1}S_{D}}%
\sum_{l=0}^{\infty }D_{l}\int_{m}^{\infty }\frac{zdz}{\sqrt{z^{2}-m^{2}}}%
\left[ 1-\frac{\xi _{1}\tilde{A}_{\alpha }B_{\alpha }}{A_{\alpha
}^2-B_{\alpha }^2(z^{2}\alpha ^{2}+\nu ^{2})}\right]   \label{paabb1} \\
&&\times \frac{\partial }{\partial \alpha }\ln \left| 1-\frac{\bar{I}_{\nu
}^{(a)}(az)}{\bar{I}_{\nu }^{(b)}(bz)}\frac{\bar{K}_{\nu }^{(b)}(bz)}{\bar{K}%
_{\nu }^{(a)}(az)}\right| ,\qquad \alpha = a,b,  \nonumber
\end{eqnarray}
where $\xi _1$ is defined in Eq.(\ref{Finperad}).
For Dirichlet and Neumann scalars the second term in the square brackets
is zero.

\section{Total Casimir energy in the region between spherical surfaces}

\label{sec:twosphtotal} \bigskip

In this section we will consider the total Casimir energy in the
region between two concentric spherical surfaces. The expression
for the unregularized VEV for the energy density can easily be
obtained from Eq.(\ref {unregWightout}) by applying the corresponding
second order operator from Eq.(\ref {vevEMT}):
\begin{equation}
\langle 0|T_{00}(x)|0\rangle =\frac{\pi ^{3}}{8a^{3}S_{D}r^{n}}%
\sum_{l=0}^{\infty }D_{l}\sum_{k=1}^{\infty }\frac{\gamma _{\nu
,k}^{3}T_{\nu }^{ab}(\gamma _{\nu ,k})}{\sqrt{\gamma _{\nu ,k}^{2}+
m^{2}a^{2}}}%
f_{\nu }^{(0)}\left[ g_{\nu }(\gamma _{\nu ,k},\gamma _{\nu ,k}r/a)\right] ,
\label{T00unreg2sph}
\end{equation}
where we use the notation (\ref{fnu0}). The total volume 
energy in the region
between the spheres is obtained by integrating this expression over $r$.
Using the standard integrals involving cylinder functions one finds
\begin{equation}
E_{a<r<b}^{{\rm (vol)}}=E_{a\leq r\leq b}+\frac{4\xi -1}{2a}
\sum_{l=0}^{\infty }D_{l}\sum_{k=1}^{\infty }
\frac{\gamma _{\nu ,k}T_{\nu }^{ab}(\gamma _{\nu ,k})}{\sqrt{
\gamma _{\nu ,k}^{2}+\mu ^{2}}}\left[ \tilde{A}_{b}B_{b}\frac{%
\bar{J}_{\nu }^{2}(\gamma _{\nu ,k})}{\bar{J}_{\nu }^{2}(b\gamma _{\nu ,k}/a)%
}-\tilde{A}_{a}B_{a}\right] ,  \label{volen2sph}
\end{equation}
where
\begin{equation}
E_{a\leq r\leq b}=\frac{1}{2}\sum_{l=0}^{\infty }D_{l}
\sum_{k=1}^{\infty }\omega _{\nu ,k},\quad \omega _{\nu ,k}=
\sqrt{\gamma _{\nu ,k}^{2}/a^2+m^{2}}, \label{toten2sph}
\end{equation}
is the total Casimir energy for the region between the plates
evaluated as a mode sum of the zero-point energies for each normal mode. 
As in the case for the region inside 
a sphere, here the energy obtained by integration
from the energy density differs from the energy (\ref{toten2sph}).
This difference can be interpreted as due to the surface energy contrbution:
\begin{equation}
E_{ab}^{{\rm (surf)}}=\frac{4\xi -1}{2a}\sum_{l=0}^{\infty
}D_{l}\sum_{k=1}^{\infty }\frac{\gamma _{\nu ,k}T_{\nu }^{ab}(\gamma _{\nu
,k})}{\sqrt{\gamma _{\nu ,k}^{2}+\mu ^{2}}}\left[ \tilde{A}_{b}B_{b}\frac{%
\bar{J}_{\nu }^{2}(\gamma _{\nu ,k})}{\bar{J}_{\nu }^{2}(b\gamma _{\nu ,k}/a)%
}-\tilde{A}_{a}B_{a}\right] .  \label{surfen2sph}
\end{equation}
To see this, recall that there is an additional contribution to the energy
density operator located on the boundaries $r=a,b$ and having the form (\ref
{surfendens}). It can be shown that by 
evaluating the corresponding VEV with the help of Eq.(\ref{fieldmodesum1ab}) 
and integrating over the region between the spheres we obtain
formula (\ref{surfen2sph}).

To derive expressions for the corresponding subtracted quantities we use the formula
\begin{equation}
\langle T_{00}^{{\rm (surf)}}(x)\rangle _{SUB}=-(2\xi -1/2)\delta
(x,\partial M)n_{r}\left( \partial _{r}\langle \varphi (r)\varphi (r^{\prime
})\rangle _{SUB}\right) _{r^{\prime }=r}  \label{Tsurfab}
\end{equation}
with the subtracted Wightman function from Eq.(\ref{regWightab1}), 
$n_{r}=1$ for $r=a+0$ and $n_{r}=-1$ for $r=b-0$. 
Using this formula for the integrated
surface energy on the surfaces $r=a+0$ (outer surface of the inner sphere)
and $r=b-0$ (inner surface of the outer sphere) one obtains
\begin{eqnarray}
E_{ab}^{{\rm (surf)}}(r) &=&n_{r}\frac{%
4\xi -1}{2\pi }\sum_{l=0}^{\infty }D_{l}\int_{m}^{\infty }\frac{zdz}{\sqrt{%
z^{2}-m^{2}}}  \label{Esurfab} \\
&\times &\left\{ \frac{\bar{I}_{\nu }^{(a)}(za)}{\bar{K}_{\nu }^{(a)}(za)}%
F_{s}[K_{\nu }(rz)]+\Omega _{a\nu }(az,bz)F_{s}[G_{\nu }^{(a)}
(az,rz)]\right\} , \nonumber
\end{eqnarray}
where we use the notation (\ref{Fs}). With the help of the alternative
representation (\ref {regWightab2}) for the Wightman function in
the region between the spheres we obtain another form of the
surface energy,
\begin{eqnarray}
E_{ab}^{{\rm (surf)}}(r) &=&n_{r}\frac{%
4\xi -1}{2\pi }\sum_{l=0}^{\infty }D_{l}\int_{m}^{\infty }\frac{zdz}{\sqrt{%
z^{2}-m^{2}}} \label{Esurfab1} \\
&\times &\left\{ \frac{\bar{K}_{\nu }^{(b)}(zb)}{\bar{I}_{\nu }^{(b)}(zb)}%
F_{s}[I_{\nu }(rz)]+\Omega _{b\nu }(az,bz)F_{s}[G_{\nu
}^{(b)}(bz,rz)]\right\} ,\;r=a+0,b-0.  \nonumber
\end{eqnarray}
The equivalence of formulas (\ref{Esurfab}) and (\ref{Esurfab1}) can
be seen directly using the identities
\begin{eqnarray}
\frac{\bar{K}_{\nu }^{(b)}(zb)}{\bar{I}_{\nu }^{(b)}(zb)}
F_{{\rm u}}[I_{\nu }(rz)]+\Omega _{b\nu }(az,bz)F_{{\rm u}}
[G_{\nu }^{(b)}(bz,rz)] &\equiv & \frac{\bar{I}%
_{\nu }^{(a)}(za)}{\bar{K}_{\nu }^{(a)}(za)}F_{{\rm u}}[K_{\nu
}(rz)] \nonumber \\
&+& \Omega _{a\nu }(az,bz)F_{{\rm u}}[G_{\nu }^{(a)}(az,rz)]
\label{idenu}
\end{eqnarray}
with u = s,v and $r=a,b$. By virtue of relations (\ref{Grelab}) the surface
energies (\ref{Esurfab}) may be presented in the form
\begin{eqnarray}
E_{ab}^{{\rm (surf)}}(r) &=& E^{{\rm (surf)}}(r)-\frac{n_{r}}{2\pi }
\left( 4\xi -1\right) \tilde{A}_{r}B_{r}\sum_{l=0}^{\infty }
D_{l}\int_{m}^{\infty }\frac{zdz}{\sqrt{z^{2}-m^{2}}}
\Omega _{r\nu }(az,bz),  \label{Esurfab2} \\
r &=&a+0,\quad b-0.  \nonumber
\end{eqnarray}
The first term on the right corresponds to the surface energy of a
single sphere and the second one is induced by the existence of
the second sphere and is finite for $a<b$. For Dirichlet and
Neumann scalars these surface energies vanish.

Integrating the vacuum energy density (\ref{compab1}) ($q=\varepsilon $)
over the region between the spheres with the help of integration formula (%
\ref{enintform}) for the corresponding volume energy, one obtains
\begin{eqnarray}
E_{ab}^{{\rm (vol)}} &=&-\frac{1}{2\pi }%
\sum_{l=0}^{\infty }D_{l}\int_{m}^{\infty }\frac{zdz}{\sqrt{z^{2}-m^{2}}}%
  \label{Evolab} \\
&\times &\left\{ \frac{\bar{I}_{\nu }^{(a)}(za)}{\bar{K}_{\nu }^{(a)}(za)}%
F_{{\rm v}}[K_{\nu }(rz)]+\Omega _{a\nu }(az,bz)F_{{\rm v}}[G_
{\nu }^{(a)}(az,rz)]\right\} _{r=a}^{r=b}.  \nonumber
\end{eqnarray}
Another form is obtained by using the energy density in the form (\ref
{compab2}) (or equivalently indentity (\ref{idenu})). 
Replacing the expression in braces at the upper limit $r=b$ in 
(\ref{Evolab}) with the help
of identity (\ref{idenu}) the expression for the volume energy in the region
$a<r<b$ can be presented in the form
\begin{eqnarray}
E_{ab}^{{\rm (vol)}} &=& E_{{\rm in}}^{{\rm (vol)}}(b)+
E_{{\rm ext}}^{{\rm (vol)}}(a)+\frac{1}{2\pi }\sum_{l=0}^{\infty
}D_{l}\int_{m}^{\infty }\frac{zdz}{\sqrt{z^{2}-m^{2}}}  \label{Evolab1}
\\
&\times &\sum_{\alpha =a,b}n_{\alpha }\Omega _{\alpha \nu }(az,bz)\left\{
(4\xi -1)\tilde{A}_{\alpha }B_{\alpha }+\left( 1-\frac{m^{2}}{z^{2}}%
\right) \left[ A_{\alpha }^{2}-\left( z^{2}\alpha ^{2}+\nu ^{2}\right)
B_{\alpha }^{2}\right] \right\} . \nonumber
\end{eqnarray}
Now the total Casimir energy in the region between the spheres may be obtained
by summing the volume and surface parts,
\begin{eqnarray}
&&E_{ab}= E_{ab}^{{\rm (vol)}}+E^{{\rm (surf)}}(a+0)+
E^{{\rm (surf)}}(b-0)=
\label{Etotab} \\
&=&-\frac{1}{2\pi }\sum_{l=0}^{\infty }D_{l}\int_{m}^{\infty }\frac{dz}{z}%
\sqrt{z^{2}-m^{2}}\left\{ \frac{\bar{I}_{\nu }^{(a)}(za)}{\bar{K}_{\nu
}^{(a)}(za)}F_{{\rm t}}[K_{\nu }(rz)]+\Omega _{a\nu }(az,bz)F_{{\rm t}}
[G_{\nu }^{(a)}(az,rz)]\right\} _{r=a}^{r=b},  \nonumber
\end{eqnarray}
where the function $\ F_{{\rm t}}[f(y)]$ is defined as
Eq.(\ref{Ftn}). Using identity (\ref{idenu}), in analogy to
(\ref{Evolab1}) (or directly summing Eqs.(\ref{Esurfab2}) and
(\ref{Evolab1})), one obtains the formula
\begin{eqnarray}
E_{ab} &=& E_{{\rm in}}(b)+E_{{\rm ext}}(a)-%
\frac{1}{2\pi }\sum_{l=0}^{\infty }D_{l}\int_{m}^{\infty }\frac{dz}{z}\sqrt{%
z^{2}-m^{2}}  \label{Etotab1} \\
&\times &\sum_{\alpha =a,b}n_{\alpha }\Omega _{\alpha \nu }(az,bz)\left[
A_{\alpha }^{2}-\left( z^{2}\alpha ^{2}+\nu ^{2}\right) B_{\alpha }^{2}%
\right] ,\;n_{b}=-1,n_{a}=1.  \nonumber
\end{eqnarray}
Note that the last (''interference'') term on the right is finite
for $a<b$ and needs no further regularization. By virtue of
relation (\ref{Omeganew}) this formula can also be presented in
the form
\begin{equation}
E_{ab} = E_{{\rm in}}(b)+E_{{\rm ext}}(a)-%
\frac{1}{2\pi }\sum_{l=0}^{\infty }D_{l}\int_{m}^{\infty } dz \sqrt{%
z^{2}-m^{2}} \frac{d}{dz}\ln \left| 1-
\frac{\bar{I}_{\nu }^{(a)}(az)}{\bar{I}_{\nu }^{(b)}(bz)}\frac{%
\bar{K}_{\nu }^{(b)}(bz)}{\bar{K}_{\nu }^{(a)}(az)}\right| .
\label{Etotab2}
\end{equation}
As we might expect the mode sum energy does not depend on the
curvature coupling parameter.

Now we turn to a system of two concentric spheres with zero
thickness. In addition to the contribution from the intermediate
region $a<r<b$ we have to include the contributions of the regions
$r<a$ and $r>b$. In these regions the vacuum densities are the
same as in the case of a single sphere. For the vacuum force
acting per unit area of the sphere with radius $r=\alpha $ one
obtains
\begin{equation}
F_{\alpha }(a,b)=F(\alpha )-n_{\alpha }p_{\alpha }(a,b,r=\alpha ),\qquad
\alpha =a,b,\;n_{a}=1,\;n_{b}=-1,  \label{force2zero}
\end{equation}
where $F(\alpha )$ is the vacuum force for a single sphere with radius $%
\alpha $ (see formula (\ref{Ftot})) and the second summand on the
right corresponds to the interaction between the spheres and is
given by Eq.(\ref{paabb}) or Eq.(\ref{paabb1}). Taking into account
formula (\ref{Etotab1}) we can see that the total vacuum energy
is equal to the sum of the vacuum energies for the separate
spheres plus an additional ''interference'' term, given by the third
summand on the right of (\ref{Etotab1}). Using expression
(\ref{Etotbar}) for a single shell Casimir energy this leads to
\begin{eqnarray}
E_{0\leq r\leq \infty }^{(ab)}&=& -\frac{1}{2\pi
}\sum_{l=0}^{\infty
}D_{l} \int_{m}^{\infty }dz\sqrt{z^{2}-m^{2}} \nonumber \\
&\times & \frac{d}{dz}\ln \left| \bar I_{\nu }^{(a)}(az)\bar
K_{\nu }^{(b)}(bz)\left[ \bar I_{\nu }^{(b)}(bz) \bar K_{\nu
}^{(a)}(az)-\bar I_{\nu }^{(a)}(az)\bar K_{\nu }^{(b)}(bz)
\right] \right| . \label{E2zero}
\end{eqnarray}
The divergences in this formula are the same as for single spheres with radii
$a$ and $b$, and can be extracted in the way described in section
\ref{sec:inen}.

\section{Conclusion} \label{sec:conclusion}

We have considered the Casimir effect for a massive scalar field
with general curvature coupling and satisfying the Robin boundary
condition on spherically symmetric boundaries in 
$D$-dimensional space. Both cases of a single sphere and two
concentric spherical surfaces are investigated.
All calculations are made at zero temperature and we assume that the
boundary conditions are frequency independent. The latter means no
dispersive effects are taken into account. The formulation of the theory
at finite temperatures can be carried out by using the standard
analyticity properties of the finite temperature Green function (see, for
instance, \cite{Kennedy}).
Unlike most
previous studies of the scalar Casimir effect here
we adopt the local approach. To obtain the expectation values for
the energy-momentum tensor we first construct the positive
frequency Wightmann function (note that the Wightmann function is
also important in considerations of the response of a particle
detector at a given state of motion). The application of the generalized
Abel-Plana formula to the mode sum over zeros of the
corresponding combinations of the cylinder functions allows us to
extract the Minkowskian part and to present the subtracted part
in terms of exponentially convergent integrals.
In this paper we consider a flat background spacetime and the subtraction
of the Minkowskian part gives finite results at any strictly interior or
exterior points in the coincidence limit.
The regularized
expectation values for the EMT are obtained by applying on the
subtracted part a certain second-order differential operator and
taking the coincidence limit. These quantities diverge as the boundary
is approached. Surface divergences are well known in quantum field
theory with boundaries and are investigated near an arbitrary shaped
smooth boundary. They lead to divergent global
quantities, such as the total energy
or vacuum forces acting on the sphere. To regularize them and to
extract numerical results we
can apply a procedure based on analytic continuation in the dimension
already used in \cite{MiltonSc,MiltonVec}
and briefly described in section \ref{sec:inen}.

  The expectation values of the EMT for the region inside a
spherical shell are given by formulas (\ref{q1in}). These
expressions are finite at interior points and diverge on the
sphere surface.  The leading term in the corresponding asymptotic
expansion is the same as for the flat boundary case and is zero
for a conformally coupled scalar. The coefficients for the
subleading asymptotic terms will depend on the boundary
curvature, Robin coefficient, and mass.
The global qunatities for the sphere interior region are
considered in section \ref{sec:inen}. Integrating the
unregularized vacuum energy density
we show that the volume energy differs from the total vacuum energy,
evaluated as the sum of the zero-point energies for the elementary
oscillators. We argue that this difference is due to the
additional surface contribution to the vacuum energy located on
the inner surface of the sphere. We give the expressions for the
subtracted volume and surface energies, Eqs.(\ref{totinenergy}) and
(\ref{surfsuben}), and for the force acting per unit area of the
sphere from inside, Eq.(\ref{Fin}). In section \ref{sec:outdens}
first we consider the scalar vacuum in the spherical layer
between two surfaces. The quantities characterizing the vacuum
outside a single sphere are obtained from this case in the limit
when the radius of the outer sphere tends to infinity.
Subtracting the parts corresponding to the space without
boundaries, for the regularized vacuum densities we derive
formulas (\ref{q1out}). These formulas differ from the ones for
the interior region by the replacements $I_{\nu }\rightarrow K_{\nu
}$, $K_{\nu }\rightarrow I_{\nu }$. We also consider the
corresponding global quantities for the outer region. As in the
interior case there is an additional surface energy, located on
the outer surface of the sphere. The decomposition of the total
outside Casimir energy into volume and surface parts is provided.
The regularized surface energy is zero for Dirichlet and Neumann
boundary conditions. In this section we further consider a
spherical shell with zero thickness and with the same Robin
coefficients on the inner and outer surfaces. The expressions for
the total and surface energies, including the contributions from
the inside
and outside regions, are obtained (formulas (\ref{Etot})
and (\ref{Etotsurf}%
)). The resulting force acting per unit area of the spherical shell
is also derived (formula (\ref{Ftot})). In section
\ref{sec:twosurfdens} we consider the expectation values for the
energy-momentum tensor in the region between
two spherical surfaces. The general case is investigated when the constants 
$A_{i}$, $B_{i}$, $i=a,b$, in the Robin boundary condition are
different for the outer and inner spheres. The expression for the
total energy in this region (including the surface parts) is
derived. The corresponding regularized Wightmann function can be
presented in different equivalent forms (Eqs.(\ref{regWightab1}),
(\ref{regWightab2}) or (\ref{intWeight})). Note that in the last
formula the ''interference'' term $\Delta W(x,x')$ defined by
Eqs.(\ref{intWeightrr}), (\ref{Wrr}) is finite in the coincidence
limit for all values $a\leq r\leq b$. The surface divergences are
contained in the two first summands on the right of
Eq.(\ref{intWeight}), corresponding to the interior and exterior
Wightmann functions of a single sphere. The components of the
vacuum EMT may be written in the forms (\ref{compab1}),
(\ref{compab2}) or (\ref{qinterf}). The additional (interaction)
force acting per unit area of the sphere with radius $r=\alpha $,
$\alpha =a,b$ due to the existence of the second sphere is given
by formula (\ref{paabb}) or by (\ref{paabb1}) and is finite for
$a<b$. In section \ref{sec:twosphtotal} we consider the total
Casimir energy and its decomposition into surface and volume
parts. First of all we show that in the general Robin case the
unregularized intermediate Casimir energy obtained by integration
of the corresponding energy density differs from the vacuum energy
evaluated as the sum of the zero-point energies for each normal
mode of frequency. As in the single sphere case this difference
may be interpreted as due to the surface energy contribution. The
corresponding subtracted quantities can be presented in the form
(\ref{Esurfab2}), where the second summand on the right is due
to the existence of the second sphere and is finite. The analogous
formulas for the integrated volume energy and total Casimir energy
in the intermediate region are in the forms (\ref{Evolab1}) and
(\ref{Etotab1}). The latter can also be presented as
Eq.(\ref{Etotab2}). As in the single sphere case the surface
contributions to the vacuum energy vanish for Dirichlet and
Neumann scalars. Then we consider a system of two concentric
spheres with zero thickness. The vacuum force acting on each
sphere can be presented as the sum of the force acting on a single
sphere plus an additional interaction force which is due to the
existence of the second sphere. The latter is given by formula
(\ref{paabb}) or (\ref{paabb1}) and is finite for $a<b$. Using
the expression for a single sphere vacuum energy the total Casimir
energy can also be presented in the form (\ref{E2zero}). For
a $D=1$ massless scalar the results given in this paper are special
cases of the general formulas derived in \cite{RomeoSah} for 
parallel plate geometry in arbitrary dimensions.

\section*{Acknowledgments}

I am grateful to Professor E. Chubaryan and Professor A. Mkrtchyan for
general encouragement and suggestions, and to A. Romeo for useful
discussions. This work was completed during my stay at
Sharif University of Technology, Tehran. It is a pleasant duty for
me to thank the Department of Physics and Professor Reza Mansouri for
kind hospitality.

\appendix

\section{Appendix: Summation formula over zeros of $C_{\protect\nu }^{ab}(%
\protect\lambda ,z)$}

Here we will consider the series over zeros of the function
\begin{equation}
C_{\nu }^{ab}(\lambda ,z)\equiv \bar{J}_{\nu }^{(a)}(z)\bar{Y}_{\nu
}^{(b)}(\eta z)-\bar{Y}_{\nu }^{(a)}(z)\bar{J}_{\nu }^{(b)}(\eta z),
\label{bescomb1}
\end{equation}
where the quantities with overbars are defined in accordance with
Eq.(\ref{barnotab}). This type of series arises in calculations of
the vacuum expectation values in confined regions with boundaries
of spherical form. To obtain a summation formula for these series
we will use the generalized Abel-Plana formula (GAPF)
\cite{Sahmat,Sahrev}. In this formula as functions $g(z)$ and
$f(z)$ let us substitute
\begin{equation}
g(z)=\frac{1}{2i}\left[ \frac{\bar{H}_{\nu }^{(1b)}(\eta z)}{\bar{H}_{\nu
}^{(1a)}(z)}+\frac{\bar{H}_{\nu }^{(2b)}(\eta z)}{\bar{H}_{\nu }^{(2a)}(z)}%
\right] \frac{h(z)}{C_{\nu }^{ab}(\eta ,z)},\quad f(z)=\frac{h(z)}{\bar{H}%
_{\nu }^{(1a)}(z)\bar{H}_{\nu }^{(2a)}(z)},  \label{gefcomb}
\end{equation}
where for definiteness we shall assume that $\eta >1$ and the 
notations $\bar{F}^{(i)},i=a,b$, are introduced in accordance with 
Eq.(\ref{barnotab}). The sum
and difference of these functions are
\begin{equation}
g(z)-(-1)^{k}f(z)=-i\frac{\bar{H}_{\nu }^{(ka)}(\lambda z)}{\bar{H}_{\nu
}^{(ka)}(z)}\frac{h(z)}{C_{\nu }^{ab}(\eta ,z)},\quad k=1,2.
\label{gefsumnew}
\end{equation}
The condition for the generalized Abel-Plana formula written in terms
of the function $h(z)$ is as follows:
\begin{equation}
|h(z)|<\varepsilon _{1}(x)e^{c_{1}|y|}\quad  |z|\rightarrow \infty
,\quad z=x+iy  , \label{cond31}
\end{equation}
where $c_{1}<2(\eta -1)$, $x^{\delta _{B_{a}0}+\delta
_{B_{b}0}-1}\varepsilon _1(x)\rightarrow 0$ for $x\rightarrow +\infty $. Let $\gamma _{\nu ,k}$ be
zeros of the function $C_{\nu }^{ab}(\eta ,z)$ in the right half plane. In
this section we will assume values of $\nu $, $A$, and $B$ 
for which all these
zeros are real and simple, and the function $\bar{H}_{\nu }^{(1a)}(z)$ ($%
\bar{H}_{\nu }^{(2a)}(z)$) has no zeros in the right half of the upper
(lower) half-plane. For real $\nu $ and $A_{a}$, $A_{b}$, $B_{a}$, $B_{b}$
the zeros $\gamma _{\nu ,k}$ are simple. To see this note that the function $%
J_{\nu }(tz)\bar{Y}_{\nu }^{(a)}(z)-Y_{\nu }(tz)\bar{J}_{\nu }^{(a)}(z)$ is
a cylinder function with respect to $t$. Using the standard result for
indefinite integrals containing the product of any two cylinder functions (see \cite
{Watson,abramowiz}) it can be seen that
\begin{equation}
\int_{1}^{\eta }{t\left[ J_{\nu }(tz)\bar{Y}_{\nu }^{(a)}(z)-Y_{\nu }(tz)%
\bar{J}_{\nu }^{(a)}(z)\right] ^{2}dt}=\frac{2}{\pi ^{2}zT_{\nu }^{ab}(\eta
,z)}\,,\quad z=\gamma _{\nu ,k},  \label{tekapositive}
\end{equation}
where we have introduced the notation
\begin{equation}
T_{\nu }^{ab}(\eta ,z)=z\left\{ \frac{\bar{J}_{\nu }^{(a)2}(z)}{\bar{J}_{\nu
}^{(b)2}(\eta z)}\left[ A_{b}^{2}+B_{b}^{2}(\eta ^{2}z^{2}-\nu ^{2})\right]
-A_{a}^{2}-B_{a}^{2}(z^{2}-\nu ^{2})\right\} ^{-1}.  \label{tekaAB}
\end{equation}
On the other hand
\begin{equation}
\frac{\partial }{\partial z}C_{\nu }^{ab}(\eta ,z)=\frac{2}{\pi T_{\nu
}^{ab}(\eta ,z)}\frac{\bar{J}_{\nu }^{(b)}(\eta z)}{\bar{J}_{\nu }^{(a)}(z)}%
\,,\quad z=\gamma _{\nu ,k}.  \label{CABderivative}
\end{equation}
Combining the last two results we deduce that for real
$\nu $, $A_{\alpha }$, $%
B_{\alpha }$ the derivative (\ref{CABderivative}) is nonzero and hence the zeros $%
z=\gamma _{\nu ,k}$ are simple. By using this it can be seen that
\begin{equation}
{\rm Res}_{z=\gamma _{\nu ,k}}g(z)=\frac{\pi }{2i}T_{\nu }^{ab}(\eta ,\gamma
_{\nu ,k})h(\gamma _{\nu ,k}).  \label{rel31}
\end{equation}
Let $h(z)$ be an analytic function for ${\rm Re}z\geq 0$ except
the poles $z_{k}$ (%
$\neq \gamma _{\nu i}$), ${\rm Re}z_{k}>0$, and with a possible branch
point at $z=0$. (For the physical application in this paper (see
section \ref{sec:twosurfdens}) the corresponding function is
analytic. However to keep the formula general we will consider
the case of a meromorphic function.) By using the GAPF, in analogy with
the derivation of the summation formula (4.13) in \cite {Sahrev}, it
can be seen that if it satisfies condition (\ref {cond31}) and
\begin{equation}
h(ze^{\pi i})=-h(z)+o(z^{-1}),\quad z\rightarrow 0,  \label{cor3cond1}
\end{equation}
and the integral
\begin{equation}
{\rm p.v.}\int_{0}^{b}{\frac{h(x)dx}{\bar{J}_{\nu }^{(a)2}(x)+\bar{Y}_{\nu
}^{(a)2}(x)}}  \label{cor2cond2}
\end{equation}
exists, then
\begin{eqnarray}
&&\lim_{b\rightarrow +\infty }\left\{ \frac{\pi ^{2}}{2}\sum_{k=1}^{m}h(%
\gamma _{\nu ,k})T_{\nu }^{ab}(\eta ,\gamma _{\nu ,k})+r_{\nu }[h(z)]-{\rm %
p.v.}\int_{0}^{b}\frac{h(x)dx}{\bar{J}_{\nu }^{(a)2}(x)+\bar{Y}_{\nu
}^{(a)2}(x)}\right\} =   \label{cor3form} \\
&=&-\frac{\pi }{2}{\rm Res}_{z=0}\left[ \frac{h(z)\bar{H}_{\nu }^{(1b)}(\eta
z)}{C_{\nu }^{ab}(\eta ,z)\bar{H}_{\nu }^{(1a)}(z)}\right] -\frac{\pi }{4}%
\int_{0}^{\infty }\frac{\bar{K}_{\nu }^{(b)}(\eta x)}{\bar{K}_{\nu }^{(a)}(x)%
}\frac{\left[ h(xe^{\pi i/2})+h(xe^{-\pi i/2})\right] dx}{\bar{K}_{\nu
}^{(a)}(x)\bar{I}_{\nu }^{(b)}(\eta x)-\bar{K}_{\nu }^{(b)}(\eta x)\bar{I}%
_{\nu }^{(a)}(x)}, \nonumber
\end{eqnarray}
where $\gamma _{\nu ,m}<b<\gamma _{\nu ,m+1}$. Here the
functional $r_{\nu }[h(z)]$ is defined as
\begin{eqnarray}
r_{\nu }[h(z)] &=&\pi \sum_{k}{\rm Res}_{{\rm Im}z_{k}=0}\left[ \frac{\bar{J}%
_{\nu }^{(a)}(z)\bar{J}_{\nu }^{(b)}(\eta z)+\bar{Y}_{\nu }^{(a)}(z)\bar{Y}%
_{\nu }^{(b)}(\eta z)}{\bar{J}_{\nu }^{(a)2}(z)+\bar{Y}_{\nu }^{(a)2}(z)}%
\frac{h(z)}{C_{\nu }^{ab}(\eta ,z)}\right] {}  \nonumber \\
&&+\pi \sum_{k,l=1,2}{\rm Res}_{(-1)^{l}{\rm Im}z_{k}<0}\left[ \frac{\bar{H}%
_{\nu }^{(lb)}(\eta z)}{\bar{H}_{\nu }^{(la)}(z)}\frac{h(z)}{C_{\nu
}^{ab}(\eta ,z)}\right] .  \label{r3}
\end{eqnarray}
In section \ref{sec:twosurfdens} formula (\ref{cor3form}) is used 
to derive the regularized vacuum energy-momentum tensor 
in the intermediate region between two spherical shells. 
Note that the corresponding functions $h(z)$ are analytic and 
hence $r_{\nu }[h(z)]=0$ for them. In the case $%
A_{a}=A_{b}$, $B_{a}=B_{b}$ from Eq.(\ref{cor3form}) one 
obtains the summation formula (4.13) in \cite{Sahrev}.

Note that Eq.(\ref{cor3form}) may be generalized 
for the functions $h(z)$ with
purely imaginary poles $\pm iy_{k}$, $y_{k}>0$, satisfying the condition
\begin{equation}
h(ze^{\pi i})=-h(z)+o\left( (z\mp iy_{k})^{-1}\right) ,\quad
z\rightarrow \pm iy_{k},  \label{cor3cond1plus2}
\end{equation}
and in the case of the existence of purely imaginary zeros $\pm
iy_{\nu ,k} $, $y_{\nu ,k}>0$ for $C_{\nu }^{ab}(\eta ,z)$. The
corresponding formula is obtained from Eq.(\ref{cor3form}) by adding
to the right hand side residue terms for $z=iy_{k}$, $iy_{\nu
,k}$ in the form
\begin{equation}
-\pi \sum_{k}{\rm Res}_{z=\eta _{k}}\left[ \frac{\bar{H}_{\nu
}^{(1b)}(\eta z)}{\bar{H}_{\nu }^{(1a)}(z)}\frac{h(z)}{C_{\nu
}^{ab}(\eta ,z)}\right] , \quad \eta _{k}=iy_{k},iy_{\nu
,k}, \label{comppoles}
\end{equation}
and taking the principal value of the integral on the right which exists 
by virtue of Eq.(\ref{cor3cond1plus2}). The arguments
here are similar to those for the Remark after Theorem 3 in \cite{Sahrev}.

\end{document}